\documentclass[final,3p,times,12pt]{elsarticle}

\usepackage{graphicx}
\usepackage{amssymb}
\usepackage{lineno}

\usepackage{gensymb}

\usepackage{amsmath}

\usepackage{bm}

\usepackage[labelformat=empty]{subfig}

\usepackage{color,soul}

\usepackage{textcomp}

\usepackage{mathtools}

\usepackage{bbold}

\usepackage{natbib}
\usepackage[colorlinks=true,linkcolor=black, citecolor=blue, urlcolor=blue]{hyperref}
\usepackage{tabularx}
\usepackage{array}
\journal{}

\begin{document}

\begin{frontmatter}


\title{Crystal plasticity model of residual stress in additive manufacturing}

%using the element elimination and reactivation method


\author[bristol]{Nicol\`{o} Grilli}
\author[engineering]{Daijun Hu}
\author[idaho]{Dewen Yushu}
\author[engineering]{Fan Chen}
\author[engineering]{Wentao Yan\corref{cor}}
\ead{mpeyanw@nus.edu.sg}

\cortext[cor]{Corresponding author}
\address[bristol]{Department of Mechanical Engineering, University of Bristol, Queen’s Building, University Walk, Bristol, UK}
\address[engineering]{Department of Mechanical Engineering, National University of Singapore, Singapore 117575, Singapore}
\address[idaho]{Computational Mechanics and Materials Department, Idaho National Laboratory, Idaho Falls, ID 83415, USA}


%\institute{N.G. and D.H. and F.C. and W.Y. (corresponding author, ) \at Department of Mechanical Engineering, National University of Singapore, Singapore 117575, Singapore \and D.Y. \at Idaho National Laboratory (INL), Idaho Falls, ID, United States}

%\date{\today}

%\maketitle

\begin{abstract}
Selective laser melting is receiving increasing interest as an additive manufacturing technique. Residual stresses induced by the large temperature gradients and inhomogeneous cooling process can favour the generation of cracks. In this work, a crystal plasticity finite element model is developed to simulate the formation of residual stresses and to understand the correlation between plastic deformation, grain orientation and residual stresses in the additive manufacturing process. The temperature profile and grain structure from thermal-fluid flow and grain growth simulations are implemented into the crystal plasticity model. An element elimination and reactivation method is proposed to model the melting and solidification and to reinitialise state variables, such as the plastic deformation, in the reactivated elements. The accuracy of this method is judged against previous method based on the stiffness degradation of liquid regions by comparing the plastic deformation as a function of time induced by thermal stresses. The method is used to investigate residual stresses parallel and perpendicular to the laser scan direction, and the correlation with the maximum Schmid factor of the grains along those directions. The magnitude of the residual stress can be predicted as a function of the depth, grain orientation and position with respect to the molten pool. 
\end{abstract}


%\PACS{62.20.fg \and 62.40.+i \and 81.40.Jj \and 64.70.dj \and 83.60.La \and 87.10.Kn}
%\subclass{74A10 \and 74B10 \and 65L60 \and 65M60 \and 65N30 \and 78M10}

\end{frontmatter}

%\pagestyle{fancy} % add AWE copyright footer on all pages
%\thispagestyle{fancy} % add AWE copyright on the first page

% deactivate line number for ArXiv submissions
%\linenumbers

\section{Introduction}
\label{intro}
Additive manufacturing (AM) is becoming increasingly popular in industries. Metal AM processes such as selective laser melting (SLM) \cite{YanSmith2016}, selective electron beam melting (SEBM) \cite{schwerdtfeger2014selective} and selective laser sintering (SLS) \cite{kruth2005binding} are layered approaches fabricating the products with complex shapes \cite{paul2014effect}. One of the bottlenecks for metal AM is the thermal stress caused by the high temperature gradients during the high-frequency heating/cooling cycles. High residual stress may result in defects \cite{li2018residual,taheri2017powder} such as the part distortion \cite{XIE2020102723}, delamination \cite{MUKHERJEE2017360} and poor fracture resistance \cite{MORI20066737,GrilliFracturePF2021,GRILLI2021100651}. The investigation of temperature history and stress evolution is therefore necessary for optimising the manufacturing parameters during the AM process. \\
Due to the large temperature gradients and high-frequency heating/cooling cycles during the manufacturing process, it is difficult to monitor the temperature and thermal stress in real time during experiments. For the residual stress measurement, the experimental approaches, such as the neutron diffraction \cite{grilli2020crystal,GrilliCOMPLAS2019}, X-ray diffraction \cite{HOCINE202030}, contour method \cite{vrancken2014residual} and semi-destructive hole drilling method \cite{mathar1934determination,martinez2019non}, are costly and sometimes limited to surface measurements. Numerical simulations are efficient to study the mechanical behaviour during the manufacturing process. The finite element method (FEM) has been widely used in simulating thermal stresses in AM \cite{ganeriwala2019evaluation}. One type of such models is based on the inherent strain method, where the deposited regions are activated with predefined strains to calculate the final deformation and residual stress of the part \cite{liang2018modified}. These models are computationally efficient, but the accuracy may be compromised due to the uncertainties of the inherent strains. Another type is thermo-mechanical coupling, where the temperature profile is calculated by solving the governing equations of heat transfer and then serves as the load for thermal stress calculation \cite{schoinochoritis2017simulation,QIAN2014312}.
%%%%%
Thermo-mechanical models are mostly constructed based on many geometrical simplifications and physical assumptions as the heat transfer part does not incorporate the thermal-fluid flow effects \cite{yan2020data}. Most of the FEM simulations failed to resolve the track morphology and molten pool evolution as all the tracks are assumed to be uniform. Therefore, those simulations are only able to predict the overall part distortion at the macroscopic scales, but cannot explain the thermal stress at the microscale. \\ 
%The anisotropic nature of metals at the microscale must be considered. The crystal plasticity finite element method (CPFEM) \cite{GRILLI2015424,TANG2021106185} complemented by grain growth models \cite{XIONG2021109410,yan2018integrated} is suitable because the anisotropy of plastic deformation is included. However, the temperature dependence of the slip mechanisms and of the elastic constants has to be included \cite{Sakano2020MD} to model the cooling process and consequent residual stress. \hl{The anisotropy of the elastic constants leads to inhomogeneous residual stresses during cooling.} \hl{These stresses are sufficient to induce some plastic deformation, which is irreversible and remains after a sample reaches room temperature.} Because of the complex thermal loading conditions induced by the laser scan \hl{at the microscale}, CPFEM simulations including a great number of grains and with an element size smaller than the grain size are necessary. \\
An emerging type of the thermal stress models is the coupling between the computational fluid dynamics (CFD) and FEM models \cite{bailey2017laser,cheon2016thermal,chen2020high}. These coupled CFD-FEM models analyse the thermal stress and deformation using the high-accuracy temperature profiles output from the CFD simulations, particularly the powder-scale thermal-fluid flow simulations \cite{YAN2018210}, instead of those from simplified analytical models \cite{WEI2021100703}. Cheon et al. \cite{cheon2016thermal} proposed a CFD-FEM framework for welding to analyse the thermal stress along a single welding line, but the model did not consider the residual stress reset due to melting. Bailey et al. \cite{bailey2017laser} implemented the temperature and surface profiles from thermal-fluid flow simulations into an FEM residual stress model to simulate the multi-track manufacturing cases of the laser directed energy deposition (DED) process. The element activation and deactivation method in ABAQUS was used to simulate the melting and solidification, which is troublesome to input command lines to activate elements of the solidified materials in the input file and frequently encounters problems on numerical convergence and numerically-induced deformation of the newly-activated elements. In our previous work \cite{chen2020high}, we simulated the multi-layer multi-track SLM process by mapping the temperature profiles from the high-fidelity thermal-fluid flow simulation into the FEM model, enabling the high resolution of the molten pool evolution and the morphology (rough surfaces and inner voids) of the FEM model. The quiet element method was used to simulate the melting and solidification, where the melted materials are assigned with null material properties. \\
There are two critical but common problems in the aforementioned thermo-mechanical models and CFD-FEM models:
\begin{enumerate}
    \item How to represent the melting and solidification while avoiding numerically-induced deformation and numerical divergence. In the previous models, two methods are commonly used: first, the element activation method \cite{stender2018thermal,montevecchi2016finite,yang2016finite,lindgren1999simulation,lindgren2001modelling} can activate elements for solidified materials but cannot deactivate elements again due to remelting, while the surface nodal temperature obtained by the interpolation between active and inactive elements \cite{michaleris2014modeling} can cause significant inaccuracy and divergence problems. Second, the quiet element method \cite{lindgren2001modelling,michaleris2014modeling,ales2018integrated}, which is essentially the same as the stiffness degradation method commonly used in fracture mechanics \cite{Miehe2010,borden2018,GrilliDuarte2018,DuarteGrilli2018,GrilliKoslowski2019}, can remove the effect of the liquid phase on the solidified region, however upon solidification the liquid phase becomes solid with initial deformation that is incompatible with the deformation of the surrounding solid region, which induces un-physical deformation and may also cause numerical divergence.
    \item None of the previous models has considered the detailed grain structures and the resultant temperature-dependent anisotropic mechanical properties at the microscale. Using simplified homogenized constitutive laws, the previous models cannot resolve the micro-scale residual stress \cite{chen2019microscale}, which can result in unique mechanical properties of the AM parts significantly differing from those by the conventional manufacturing \cite{yan2020MRL}.
\end{enumerate}
In this paper, a crystal plasticity finite element model (CPFEM) \cite{GRILLI2015424,GrilliDIC2020} is developed, incorporating the temperature profile from the thermal-fluid flow simulation \cite{chen2020high} and grain structures from the grain growth simulation \cite{YangMin2021,yan2018integrated} and experimental data \cite{chen2019microscale}, as to be described in Section \ref{sec:cpmodel}. Particularly, an element elimination and reactivation method is proposed to model the melting and solidification. More importantly, the algorithm can reinitialise the state variables, as tested in a simple case in Section \ref{sec:testcase}, such as the plastic deformation of the solidifying elements, which is critical to avoid the pre-existing plastic deformation just after solidification. Simulation cases of SLM of 316 stainless steel are conducted as described in Section \ref{sec:simulationslm}. The proposed element elimination and reactivation method and the commonly used stiffness degradation method are compared in terms of prediction accuracy, convergence and computational efficiency. Based on the simulation results, the formation of residual stresses in different grains and the correlation between plastic deformation, depth and residual stress are discussed in Section \ref{sec:discussion}.

%The temperature field predicted by CFD simulations is used for a grain growth model. The grain structure found is used for CPFEM simulations, together with the temperature field. A comparison with the residual stiffness method is carried out to understand the differences in the prediction of the plastic deformation. 

%Reinitialisation of the displacement field in activating elements allows to consider the thermal expansion of the substrate. 

%This method will be called the ``element elimination and reactivation method'' in the following. 

%Section \ref{sec:cpmodel} includes the crystal plasticity material model and the two methods to reproduce material deposition. Section \ref{sec:testcase} reports a test case in which the element elimination and reactivation method is applied on a simple geometry under tension.

\section{Material model}
\label{sec:cpmodel}

\subsection{Crystal plasticity framework}
\label{sec:crystalplastframework}

A crystal plasticity material model including thermal response is used for this work.
The deformation gradient $\boldsymbol{F}$ can be decomposed into elastic, thermal and plastic parts \cite{asaro1977strain,kalidindi1998incorporation,vujovsevic2002finite,GRILLI2020104061}:
\begin{equation}
\boldsymbol{F}=\boldsymbol{F}_e \boldsymbol{F}_{th} \boldsymbol{F}_p \ ,
\label{eqn:elastoplasticdecomposition}
\end{equation}
where $\boldsymbol{F}_e$ is the elastic part including the stretching and rigid body rotation, $\boldsymbol{F}_{th}$ is thermal part and $\boldsymbol{F}_p$ is the irreversible plastic deformation, which evolves according to \cite{roters2010overview,ROTERS2019420}:
\begin{equation}
    \boldsymbol{L}_p = \dot{\boldsymbol{F}}_p \boldsymbol{F}_p^{-1} = \sum_{\alpha=1}^n \dot{\gamma}^{\alpha}\boldsymbol{m}^{\alpha} \otimes\boldsymbol{n}^{\alpha} \ ,
\end{equation}
where $\dot{\gamma}^{\alpha} $ is the shear rate of slip system $\alpha$. The vectors $\boldsymbol{m}^{\alpha}$ and $\boldsymbol{n}^{\alpha}$ are unit vectors which describe the slip direction and normal of slip plane $\alpha$, respectively. $n$ is the number of slip systems in active state. \\
The plastic strain rate on slip system $\alpha$ of face-centered cubic metals can be calculated by a power-law equation \cite{rice1971inelastic,peirce1983material}:
\begin{equation}
\label{eqn:gammadot}
\dot{\gamma}^{\alpha}=\dot{\gamma}_0 \left | \frac{\tau^{\alpha}}{\tau_c^{\alpha}(T)} \right |^{\frac{1}{m}} \rm{sgn}(\tau^{\alpha}) \ ,
\end{equation}
where $\tau^{\alpha}$ is the resolved shear stress, $\tau_c^{\alpha}$ is the temperature dependent critical resolved shear stress (CRSS), $\dot{\gamma}_0$ and $m$ are constants, which indicate the reference strain rate and rate sensitivity of slip, respectively. \\
An exponential model calibrated by experiments can be used to calculate the CRSS change with temperature decrease of each slip system \cite{daymond2006elastoplastic,ORNLpaper2020}:
\begin{equation}
\tau_c^{\alpha}(T)= \left ( k_A+k_{B}\cdot{\rm exp} [-k_{C}(T-T_0)] \right ) \cdot\tau_c^{\alpha}(T_0) \ ,
\end{equation} 
where $\tau_c^{\alpha}(T)$ is the CRSS of slip system $\alpha$ at temperature $T$, $T_0$ is the reference temperature, $k_A$, $k_B$ and $k_C$ are constants. For slip system $\alpha$, its hardening behaviour is influenced by other slip systems $\beta$ as \cite{kalidindi1992approximate}:

\begin{equation}
\dot{\tau}_c^{\alpha} = \sum_{\alpha=1}^{n}h_{\alpha\beta} \left | \dot{\gamma}^{\beta} \right | \ ,
\end{equation}
where $h_{\alpha\beta}$ is the hardening matrix given by \cite{Diehl2017,IRASTORZALANDA2016184}:
\begin{equation}
h_{\alpha\beta} = q_{\alpha\beta} \left [ h_0 \left ( 1-\frac{{\tau_c^{\alpha}}}{\tau_\textrm{sat}} \right )^{a} \right ] \ ,
\end{equation}
where $q_{\alpha\beta}$ is a hardening coefficient matrix for latent hardening, $h_0$ is an initial hardening term, $\tau_{sat}$ is the saturation slip resistance and $a$ is a constant.
The parameters of the crystal plasticity model are calibrated using the experimental data in \cite{chen2019microscale}. \\
For the thermo-mechanical response during the laser scan process, the Green-Lagrange strain tensor can be expressed by the equation composed of thermal and elastic deformation gradients as:
\begin{equation}
\boldsymbol{E}_e = \frac{1}{2} \left ( \boldsymbol{F}_\textrm{th}^T\boldsymbol{F}_{e}^T\boldsymbol{F}_{e}\boldsymbol{F}_\textrm{th}-\boldsymbol{I} \right ) \ ,
\end{equation} 
where $\boldsymbol{I}$ is the identity matrix. To describe how the size of the substance changes with the temperature, the linear thermal expansion coefficient $\alpha_l$ of 316L SS is introduced \cite{callen1998thermodynamics}. The volumetric thermal expansion coefficient can be expressed by $\alpha_v=3\alpha_l$ \cite{Dubrovinsky2002}. Since the thermal expansion coefficient of 316L stainless steel varies slightly with temperature  \cite{yadroitsev2015evaluation,grimvall1999thermophysical}, and the temperature dependence is given by fitting the experimental data into a linear equation \cite{jiang2012using}:
\begin{equation}
\alpha_v=\alpha_0+\alpha_1(T-T_0) \ ,
\label{eqn:linearTexpansion}
\end{equation} 
where $\alpha_0$ denotes the thermal expansion coefficient at room temperature and $\alpha_1$ is the linear coefficient of the temperature dependence of $\alpha_v$. The infinitesimal volumetric change $dV$ from ${T_0}$ to ${T}$ can be described by \cite{vujovsevic2002finite}:

\begin{equation}
\alpha_{v}\mathrm{d}T= \frac{\mathrm{d}V}{V} \ ,
\label{eqn:infinitesimalTexpansion}
\end{equation} 
where $V$ is the volume of the region and $T_0$ is the reference temperature (room temperature). Substituting equation (\ref{eqn:linearTexpansion}) into equation (\ref{eqn:infinitesimalTexpansion}), the thermal eigenstrain tensor $\boldsymbol{\alpha}$ can be expressed by \cite{grilli2018effect}:
\begin{equation}
\boldsymbol{\alpha}=\frac{1}{2}\left[{\rm exp}\left(\frac{1}{3}\alpha_{1}(T-T_{0})^2+\frac{2}{3}\alpha_{0}(T-T_0) \right) - 1 \right] \boldsymbol{I} \ .
\end{equation}
The 2nd Piola-Kirchhoff stress can be calculated by:
\begin{equation}
\boldsymbol{S}=\mathbb{C}(\boldsymbol{E_e}-\boldsymbol{\alpha}) \ ,
\label{eqn:pkstress}
\end{equation}
where $\mathbb{C}$ is the rank four elasticity tensor of 316L stainless steel \cite{Horiuci316SS,GRILLI2018104}. The elasticity tensor of the solid is temperature dependent \cite{daymond2006elastoplastic}, which can be described by a linear function for the components:
\begin{equation}
\mathbb{C}_{ij}^{\rm{solid}}(T)=\mathbb{C}_{ij}^{\rm{solid}}(T_0) + \frac{\mathrm{d}\mathbb{C}_{ij}^{\rm{solid}}}{\mathrm{d}T}(T-T_0) \ ,
\label{eqn:cijtempdep}
\end{equation}
where $\mathbb{C}_{ij}^{\rm{solid}}(T_0)$ is the elasticity tensor component at room temperature, and the derivative term indicates the change of the components caused by temperature change. Phase change can be modelled by changing the stiffness tensor $\mathbb{C}$, as to be explained in section \ref{sec:residualstiffmethod}. \\
Since positive strain represents tension and negative strain represents compression, the same applies to the stress tensor components. In the following, compression will appear as a negative stress component while tension will appear as a positive stress component. \\
%By Chen Fan%
The temperature $T$ is obtained from thermal-fluid flow simulations \cite{chen2020high}. It should be mentioned that the temperature mapping from thermal-fluid to mechanical simulation is one-way, and the heat transfer analysis is merely conducted in the high-fidelity thermal-fluid flow simulation, from which the thermal information are extracted for the crystal plasticity calculation. Thus, deactivation of the melted parts will not affect the temperature distribution in the representative volume. %By Chen Fan%

\subsection{Residual stiffness method}
\label{sec:residualstiffmethod}

\begin{figure}[!htb]
\centering
\includegraphics[width=0.5\textwidth]{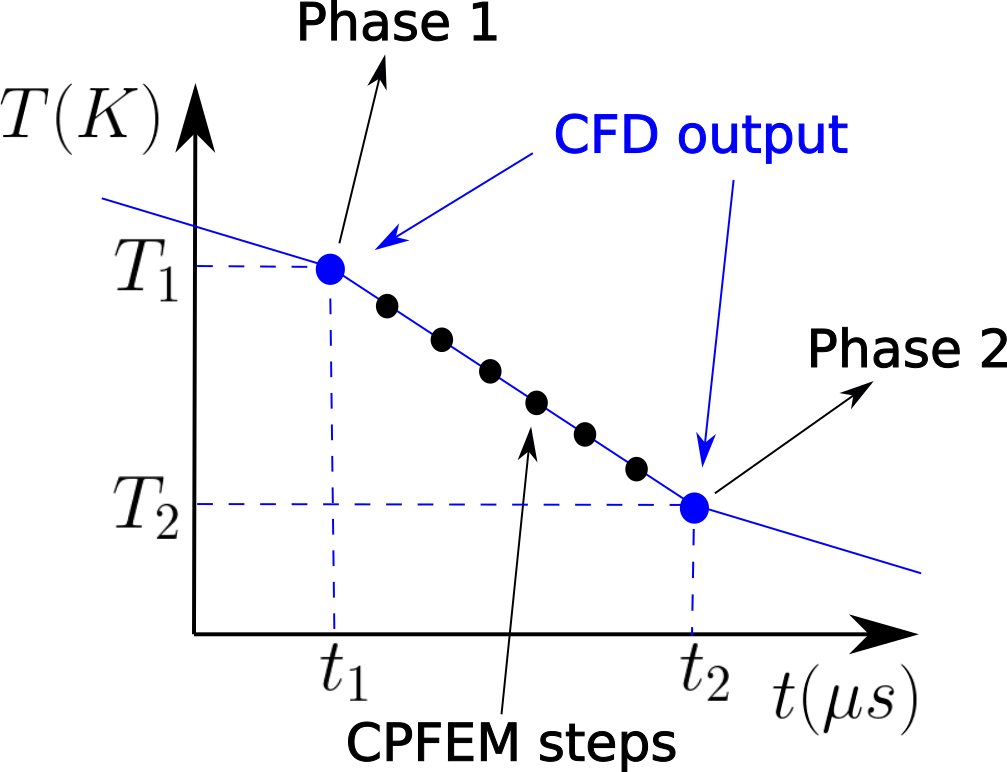}
\caption{\label{fig:cfdsteps} Schematic of the phase identification algorithm to implement the residual stiffness method for liquid and gas phases.}
% /home/nicolo/projects/c_pfor_am_Test/CFDGrainGrowth/Slides/Images/TimeInterpolationPhase.png
\end{figure}

During the additive manufacturing process, the phase transitions cause significant change in the mechanical properties, which should be taken into consideration.
Since in our simulation the temperature profiles are obtained by mapping and interpolating the thermal-fluid flow simulation results to the finite element model in time and space \cite{YanLin2018,chen2020high}, there are liquid phase and gas phase existing in the model, which represent the melted material and gas above molten pool, respectively. Therefore, a residual stiffness method based on the temperature field is applied here to prevent sudden changes in mechanical properties caused by the temperature dependence of stiffness and avoid the convergence problem of simulation. 
The specific method is to set a temperature range between the melting point $T_m$ of 316L stainless steel and the gas temperature $T_g$ set in the thermal-fluid flow simulation. In this temperature range, the stiffness tensor of the solid material is $\mathbb{C}_{ij}^{\rm{solid}}(T)$, which conforms to equation (\ref{eqn:cijtempdep}). \\
The temperature from the thermal-fluid flow simulation is recorded every $45$ $\mu$s and these time points are called the ``CFD output'' or ``temperature output''. The CPFEM simulations have smaller time steps, therefore an interpolation is necessary. For instance, in the time interval $\left [ t_1, t_2 \right ]$ the temperature is interpolated linearly from $T_1$ to $T_2$, as shown in Figure \ref{fig:cfdsteps}. \\ 
A phase identification algorithm has been implemented: if the temperature is higher than the melting temperature $T_m$ or lower than the gas temperature $T_g$, it means that the material is in liquid or gas phase. This algorithm transforms the stiffness matrix $\mathbb{C}_{ij}$ linearly with time between those of phase $1$ at $t_1$ and phase $2$ at $t_2$ in Figure \ref{fig:cfdsteps}. The formulas to find the stiffness tensor are listed in Table.\ref{tab:cfdsteps}. A residual stiffness coefficient $q_r$ is introduced to describe the residual stiffness tensor of the material:
\begin{equation}
\mathbb{C}_{ij}^{\rm{residual}}=q_{r}\cdot\mathbb{C}_{ij}^{\rm{solid}}(T_0) \ .
\label{eqn:residualstiff}
\end{equation}
The stiffness tensor $\mathbb{C}_{ij}$ in Table \ref{tab:cfdsteps} is used in equation (\ref{eqn:pkstress}) and the interpolation avoids discontinuities in  the stiffness tensor that would create convergence problems. \\
\begin{table}[!htb]
\normalsize
\caption{Time evolution of the stiffness tensor depending on the phase state at time $t_1$ and $t_2$.}
\label{tab:cfdsteps}
\centering
\def\arraystretch{1.5}
\begin{tabular}{|c|c|c|}
\hline
Phase 1 & Phase 2 & Stiffness \\
\hline
Solid & Solid & $\mathbb{C}_{ij} = \mathbb{C}_{ij}^\mathrm{solid} \left ( T \right )$ \\
\hline 
Solid & Gas/Liquid & $\mathbb{C}_{ij} = \frac{(t_2-t)}{(t_2-t_1)}\mathbb{C}_{ij}^\mathrm{solid} \left ( T_1 \right )
+ \frac{(t-t_1)}{(t_2-t_1)}\mathbb{C}_{ij}^\mathrm{residual}$ \\
\hline  
Gas/Liquid & Solid & $\mathbb{C}_{ij} = \frac{(t_2-t)}{(t_2-t_1)}\mathbb{C}_{ij}^\mathrm{residual}
+ \frac{(t-t_1)}{(t_2-t_1)}\mathbb{C}_{ij}^\mathrm{solid} \left ( T_2 \right )$ \\
\hline
Gas/Liquid & Gas/Liquid & $\mathbb{C}_{ij} = \mathbb{C}_{ij}^\mathrm{residual}$ \\
\hline
%$\mathbb{C}_{ij} = \mathbb{C}_{ij}^\textrm{residual}$
\end{tabular}
\end{table}
The equations above are implemented into the MOOSE finite element framework \cite{permann2020moose}, where a Newton-Raphson approach to find the Cauchy stress tensor $\boldsymbol{\sigma}$ is applied. To obtain the 2nd Piola-Kirchhoff stress in equation (\ref{eqn:pkstress}), a function $\boldsymbol{\psi}=\boldsymbol{S}-\mathbb{C}(\boldsymbol{E_e}-\boldsymbol{\alpha})$ is defined, the absolute value of which must be minimised. An iterative return mapping algorithm, in which the stress $\boldsymbol{S}$ is the variable is implemented. The stress increment during the return mapping algorithm can be written as \cite{adhikary2016robust,chockalingam2013crystal}: 
\begin{equation}
\boldsymbol{S}_{i+1}=\boldsymbol{S}_{i}-\left( \frac{\partial\boldsymbol{\psi}}{\partial\boldsymbol{S}} \right)^{-1}\cdot\boldsymbol{\psi} \ .
\end{equation}
where $i$ and $i+1$ indicate two subsequent iterations. The Cauchy stress $\boldsymbol{\sigma}$ is obtained by \cite{lee1969elastic}:
\begin{equation}
\boldsymbol{\sigma}=\frac{1}{\rm{det}(\boldsymbol{F_e})}\cdot\boldsymbol{F_e}\cdot\boldsymbol{S}\cdot\boldsymbol{F_{e}^T} \ .
\end{equation}
The Cauchy stress is used by the finite element solver to find the displacement field $\mathbf{u}$ that leads to stress equilibrium. \\
The parameters used in the simulations and the elastic constants of 316L stainless steel in Voigt notation are listed in Table \ref{tab:modelparameters}.

\begin{table}[!htb]
\normalsize
\caption{Material and model parameters used in the simulations \cite{daymond2006elastoplastic,chen2019microscale,jiang2012using,irastorza2017effect,Clausen1998}}
%}
\label{tab:modelparameters}
\centering
\begin{tabular}{ll}
\hline\noalign{\smallskip}
Reference plastic strain rate ($\dot{\gamma}_0$) & $1 \times  10^{-7}$ s$^{-1}$  \\
Plastic strain exponent constant ($m$) & 0.1 \\
Hardening matrix ($h_{\alpha\beta}$) & 3839 MPa  \\
Hardening exponent ($a$) & 2.5  \\
Reference temperature ($T_0$) & 303 K \\
Melting temperature ($T_m$) & 1648.15 K  \\
Gas temperature ($T_g$) & 298.1 K \\
Critical resolved shear stress at $T_0$ ($\tau_c^{\alpha}$) & 213 MPa  \\
Saturation slip resistance ($\tau_{sat}$) & 302 MPa  \\
Volumetric thermal expansion coefficient at $T_0$ ($\alpha_0$) & 44.73$\times 10^{-6}$ K$^{-1}$ \\
Derivative of volumetric thermal expansion coefficient ($\alpha_1$) & 0.01011$\times 10^{-6}$ K$^{-2}$ \\
Temperature dependence of CRSS ($k_A$) & 0.53 \\
Temperature dependence of CRSS ($k_B$) & 0.47 \\
Constant in temperature dependence eq.(4) of CRSS ($k_C$) & 0.008 \\
Elastic constants at $T=T_0$ ($\mathbb{C}_{11}$) & 204.6 GPa \\
Elastic constants at $T=T_0$ ($\mathbb{C}_{12}$) & 137.7 GPa \\
Elastic constants at $T=T_0$ ($\mathbb{C}_{44}$) & 126.2 GPa \\
Derivative of elastic constant ($\mathrm{d}\mathbb{C}_{11}/{\mathrm{d}T}$) & -90.33 MPa K$^{-1}$ \\
Derivative of elastic constant ($\mathrm{d}\mathbb{C}_{12}/{\mathrm{d}T}$) & -45.10 MPa K$^{-1}$ \\
Derivative of elastic constant ($\mathrm{d}\mathbb{C}_{44}/{\mathrm{d}T}$) & -51.78 MPa K$^{-1}$ \\
Residual stiffness coefficient ($q_r$) & 0.01 and 0.001 \\
Laser scan speed & 700 mm/s \\
Laser power & 150 W \\
\noalign{\smallskip}\hline
\end{tabular}
%}
\end{table}

\subsection{Implementation of the element elimination and reactivation method}
\label{sec:elemelimreactimpl}

% This section describes the implementation of the element elimination and reactivation method in the MOOSE framework \cite{permann2020moose}.

To model the melting and solidification of material in the SLM process, an element elimination and reactivation approach is employed. This approach is based on a moving subdomain paradigm that is originally developed \cite{Yushu2021ElemElim}, which is based on MOOSE’s subdomain restricted system \cite{permann2020moose}. Specifically, the physical domain is divided into an active subdomain ($\Omega_a$) and an inactive subdomain ($\Omega_i$). The thermo-mechanical finite element computation is defined and carried out in $\Omega_a$, while $\Omega_i$ only carries the geometric information. In other words, the elements in the inactive domain, or inactive elements, do not contribute to the equation system that calculates stress equilibrium. \\
The active and inactive subdomains are determined only by the prescribed temperature field. At a typical FEM time step, any active element $\mathcal{T} \in \Omega_a $ with an average temperature that is above the melting temperature (i.e., $T_{\text{avg}} > T_m$) or below the gas temperature (i.e., $T_{\text{avg}} < T_g$) is moved from the active subdomain to the inactive one, i.e.:
\begin{equation}
\Omega_a \rightarrow \Omega_a \backslash \mathcal{T}, \quad \Omega_i \rightarrow \Omega_i \cup \mathcal{T}. 
\end{equation}
 This process mimics the solid material being eliminated from the physical domain, which happens when the material is melted around the laser beam, and is therefore referred to as element elimination.
Similarly, when the average temperature of one inactive element $\mathcal{T} \in \Omega_i $ falls to the solid temperature range (i.e., $T_g \leq T_{\text{avg}} \leq T_m$), it is moved to the active subdomain, i.e.: 
\begin{equation}
\Omega_a \rightarrow \Omega_a \cup \mathcal{T}, \quad \Omega_i \rightarrow \Omega_i \backslash \mathcal{T}. 
\end{equation}
This mimics the process of material being added back to the solid physical domain, which happens when the material solidifies behind the laser beam,  thus is referred to as element reactivation. 

After elements being transferred between the two subdomains, several tasks need to be accomplished before the subsequent computation. First, the boundary information is updated based on the coverage of the new subdomains. Here, the subdomain boundaries are updated for both $\Omega_a $ and $\Omega_i$. The boundary conditions are projected to the updated boundaries correspondingly.
Second, the initial conditions of the displacement variables and the material properties that define the crystal plasticity behaviour, such as the plastic deformation (see Section \ref{sec:crystalplastframework}), are projected to the quadrature points of the reactivated elements. Projection of the initial condition indicates that the displacement, strain, and stress calculations are reset to zero. This allows for the restart of the plastic deformation calculation during the simulation, which cannot be implemented easily in the stiffness degradation method \cite{Duarte2019APS}. Indeed, if the elements are not eliminated and reactivated, they can still carry a certain amount of load. In the stiffness degradation method, a sudden change of the plastic deformation due to melting and solidification causes convergence problems because an elastic deformation jump, and consequent stress discontinuity, occur. \\
Finally, the element elimination and reactivation method resolves the issue that is brought by large deformation from the reactivated elements if they are set to carry the displacement history upon reactivation.

%The code for the material model described in section \ref{sec:cpmodel} is available in the following repository \cite{cpforam2021}.

\section{Test case: element elimination during tension}
\label{sec:testcase}

\begin{figure}[!htb]
\centering
\subfloat[(a)]{\includegraphics[width=0.35\textwidth]{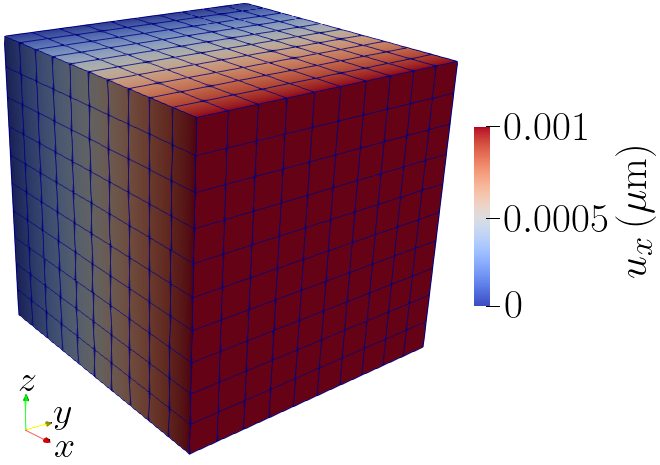}}
%\hspace{0.1cm}
\subfloat[(b)]{\includegraphics[width=0.35\textwidth]{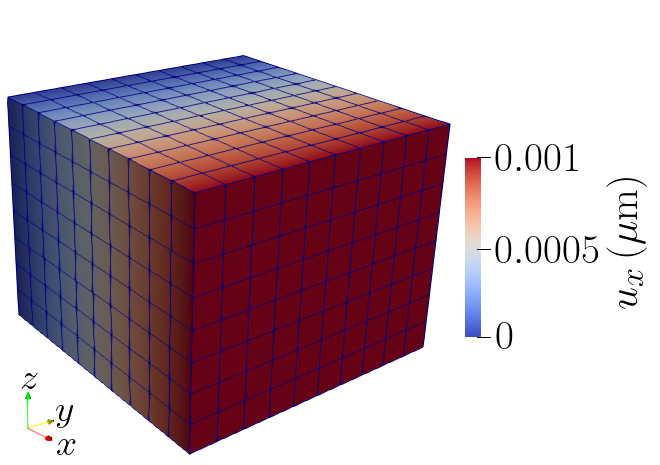}}
\subfloat[(c)]{\includegraphics[width=0.35\textwidth]{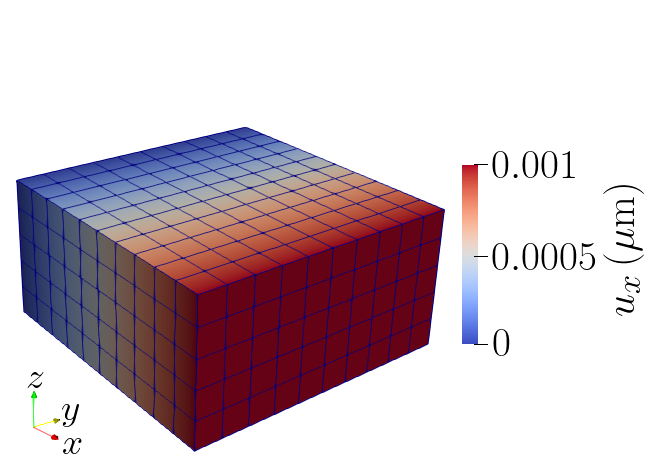}}
\newline
\subfloat[(d)]{\includegraphics[width=0.38\textwidth]{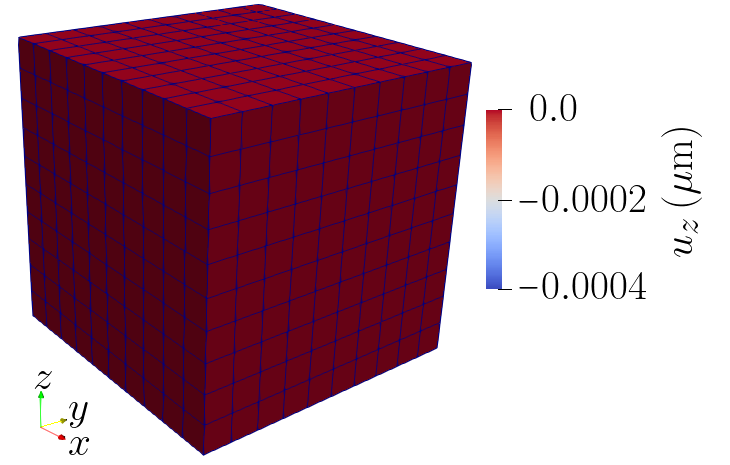}}
\subfloat[(e)]{\includegraphics[width=0.38\textwidth]{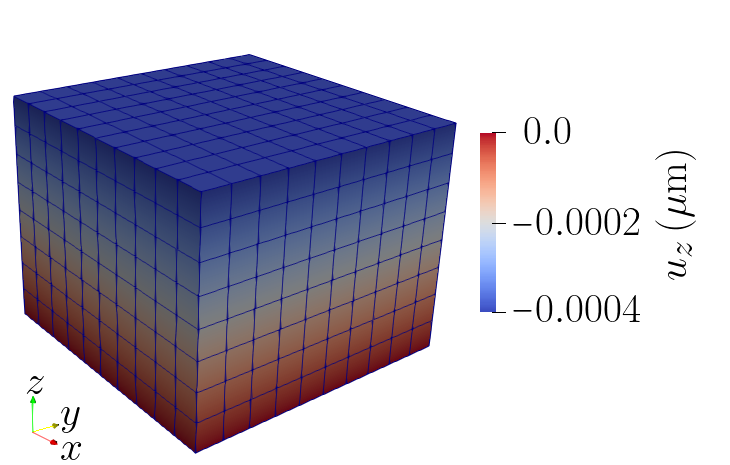}}
\subfloat[(f)]{\includegraphics[width=0.38\textwidth]{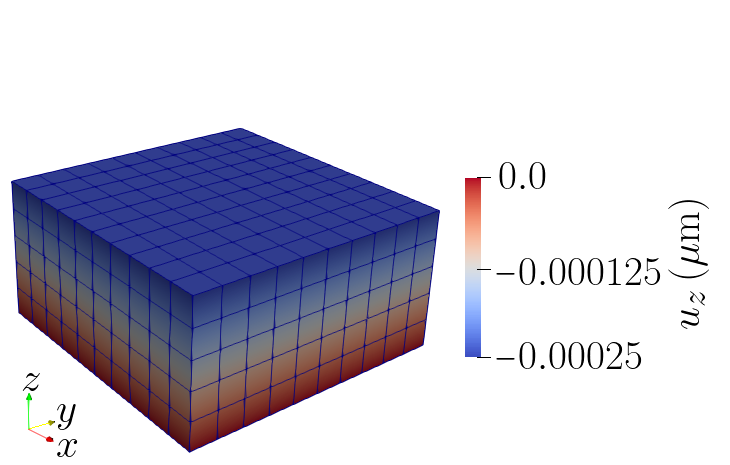}}
\newline
\subfloat[(g)]{\includegraphics[width=0.38\textwidth]{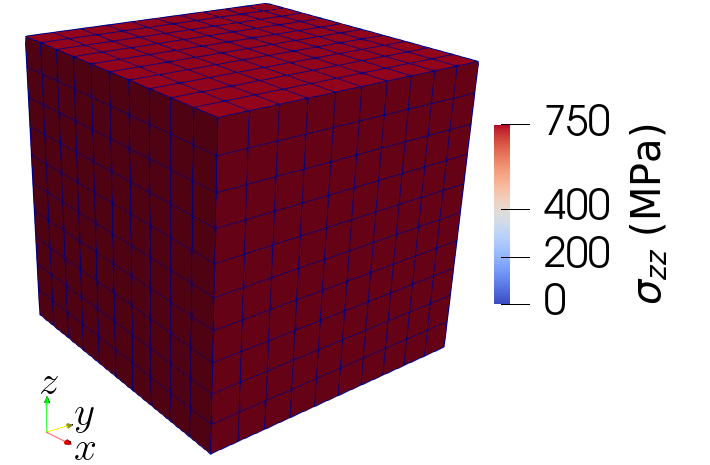}}
\subfloat[(h)]{\includegraphics[width=0.38\textwidth]{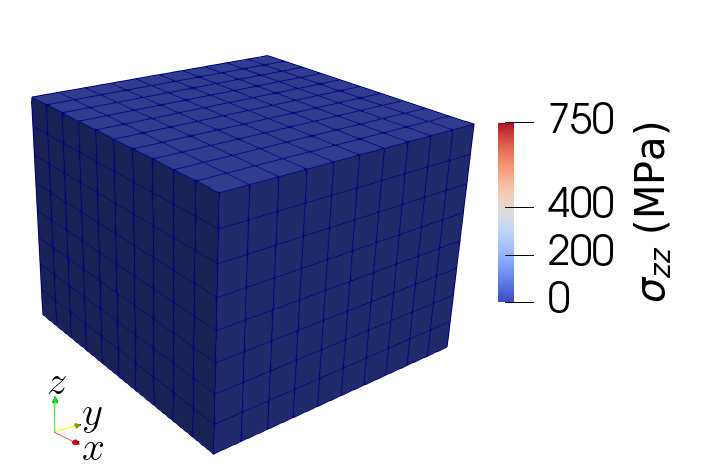}}
\subfloat[(i)]{\includegraphics[width=0.38\textwidth]{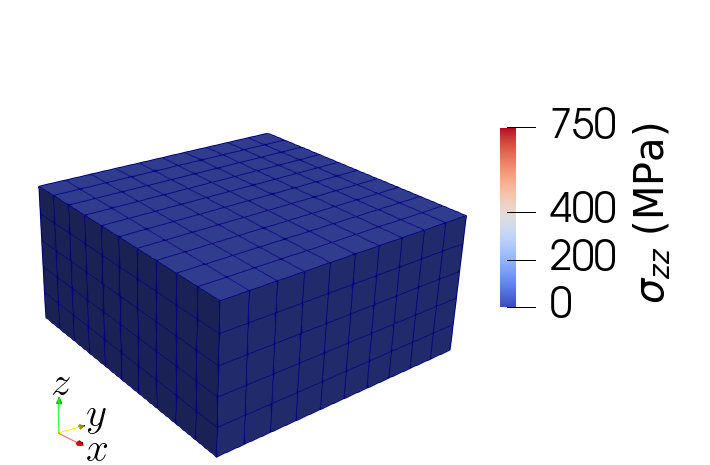}}
\caption{\label{fig:methodtest} Test case for the element elimination and reactivation method. (a),(b),(c) Firstly, a displacement $\mathbf{u}_x$ is applied and elements are eliminated. (d) The lateral displacement starts at zero, (e)-(f) then lateral contraction appears due to the newly created free surface perpendicular to the $z$ axis. For the same reason, (g) the lateral stress $\boldsymbol{\sigma}_{zz}$ starts at 750 MPa, (h)-(i) then $\boldsymbol{\sigma}_{zz}$ becomes zero.}
% /home/nicolo/projects/c_pfor_am_Test/ElementAddDelete/Block2DispTD/
%/home/nicolo/projects/c_pfor_am_Test/ElementAddDelete/DispX/View.pvcc
%/home/nicolo/projects/c_pfor_am_Test/ElementAddDelete/Block2DispTD/DispZTime5Scale2.png
%/home/nicolo/projects/c_pfor_am_Test/ElementAddDelete/Block2DispTD/SigmaZZTime5Scale2.png
%/home/nicolo/projects/c_pfor_am_Test/ElementAddDelete/Block2DispTD/SigmaZZTime20Scale2.png
%/home/nicolo/projects/c_pfor_am_Test/ElementAddDelete/Block2DispTD/SigmaZZTime90Scale2.png
\end{figure}

The element elimination and reactivation method is tested on a simple example geometry, as shown in Figure \ref{fig:methodtest}. The aim of this test is to demonstrate that inactive elements do not contribute to the stress equilibrium calculations, i.e. they do not exert stress on the active elements, and to prove that boundary conditions are implemented correctly on the continuously changing free surface. \\
A cubic representative volume is constituted of 10 $\times$ 10 $\times$ 10 elements and has a size of 1 $\mu$m $\times$ 1 $\mu$m $\times$ 1 $\mu$m. Boundary condition $\mathbf{u}_x = 0$ is applied on the surface $x = 0$, $\mathbf{u}_y = 0$ on $y=0$ and $y = 1$ $\mu$m, $\mathbf{u}_z = 0$ on $z=0$ and $z = 1$ $\mu$m. Displacement boundary condition is imposed on the surface $x = 1$ $\mu$m. $\mathbf{u}_x$ is increased up to $0.001$ $\mu$m on that surface before the element elimination algorithm starts. \\
The element elimination algorithm is based on a prescribed temperature field, which is chosen in such a way that elements from the surface $z = 1.0$ $\mu$m are progressively removed, down to the surface $z = 0.5$ $\mu$m. The displacement $\mathbf{u}_x$ on the surface $x = 1$ $\mu$m is kept constant during element elimination, as shown in Figure \ref{fig:methodtest}(a)-(c), therefore $\boldsymbol{\varepsilon}_{xx}$ remains constant at the value of $0.001$. \\ 
Plasticity and thermal eigenstrain are not activated in this simulation case, and the elastic constants are set as $\mathbb{C}_{11} = 1.5$ GPa and $\mathbb{C}_{12} = 0.75$ GPa. $\boldsymbol{\varepsilon}_{yy}$ and the off diagonal strain components are zero. When no element elimination has taken place, the boundary condition $\mathbf{u}_z = 0$ on the surface $z=1$ $\mu$m holds, therefore $\boldsymbol{\varepsilon}_{zz} = 0$ and $\mathbf{u}_z = 0$ everywhere in the representative volume, as shown in Figure \ref{fig:methodtest}(d). This lateral constraint is reflected by the large positive value of the stress component $\boldsymbol{\sigma}_{zz}$ in Figure \ref{fig:methodtest}(g). Therefore, a tensile stress is present on that surface. \\
After the element elimination algorithm starts, the top surface perpendicular to the $z$ axis becomes unconstrained. Therefore, lateral contraction takes place because of the constant displacement along the $x$ axis. The lateral contraction can be calculated as follows: $\boldsymbol{\varepsilon}_{zz} = - \left ( \mathbb{C}_{12} / \mathbb{C}_{11} \right ) \boldsymbol{\varepsilon}_{xx} = -0.0005$. For example, in Figure \ref{fig:methodtest}(e), 8 elements are still present along the $z$ axis, therefore the lateral displacement on the top surface is given by $\mathbf{u}_z = \boldsymbol{\varepsilon}_{zz} \cdot 0.8$ $\mu$m $= -0.0004$ $\mu$m. The same holds in Figure \ref{fig:methodtest}(f), in which the lateral displacement on the top surface is $\mathbf{u}_z = -0.00025$ $\mu$m because 5 elements are still present along the $z$ axis. \\
As shown in Figures \ref{fig:methodtest} (h)-(i), there is no lateral stress $\boldsymbol{\sigma}_{zz}$ after the element elimination algorithm starts. This proves that the method developed creates new free surfaces in the representative volume and the eliminated elements do not apply stress on the active elements.

\section{Simulation of selective laser melting}
\label{sec:simulationslm}

\subsection{Simulation setup}
\label{sec:simulationsetup}

\begin{figure}[!htb]
\centering
\includegraphics[width=0.6\textwidth]{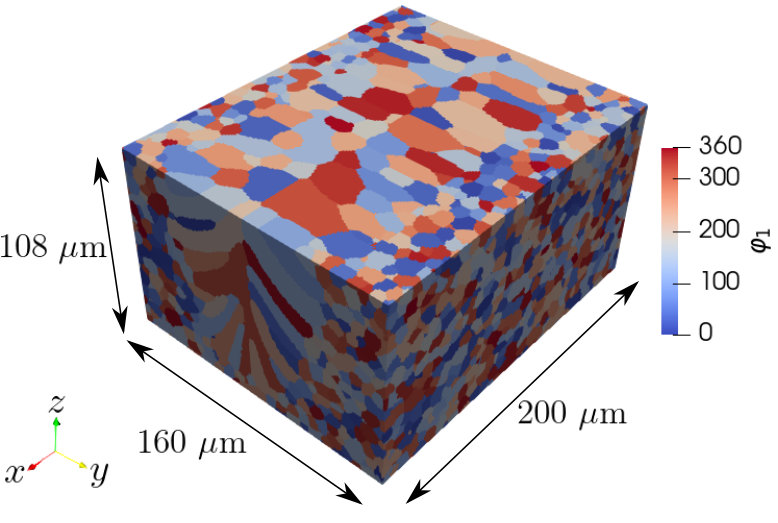}
\caption{\label{fig:grains} Representative volume for the selective laser melting simulations and grain structure implemented from the phase field simulation for the crystal plasticity model.}
% /home/nicolo/projects/c_pfor_am_Test/SpreadAndCFD/Grains1um/Grains.png
%/home/nicolo/projects/c_pfor_am_Test/SpreadAndCFD/Residualstress4um/Case2DegradEigenBFI3_out/Case2DegradEigenBFI3_out.e
% /home/nicolo/projects/c_pfor_am_Test/SpreadAndCFD/Grains1um/GrainsLength3.png
\end{figure}

\begin{figure}[!htb]
\centering
% Time 99 us
\subfloat[(a)]{\includegraphics[height=0.3\textwidth]{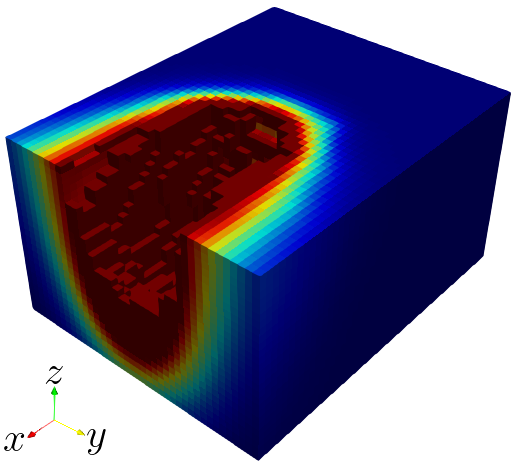}}
% Time 240 us
\subfloat[(b)]{\includegraphics[height=0.3\textwidth]{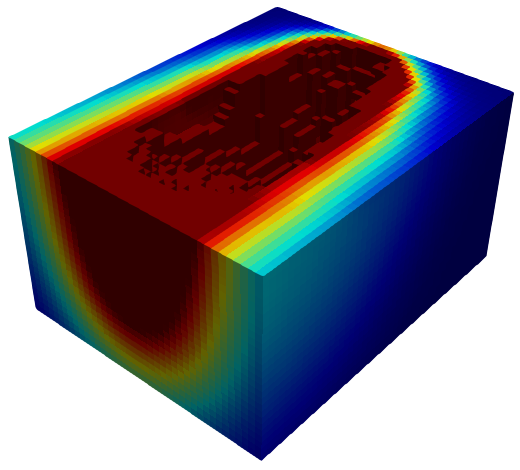}}
% Time 472 us
\subfloat[(c)]{\includegraphics[height=0.3\textwidth]{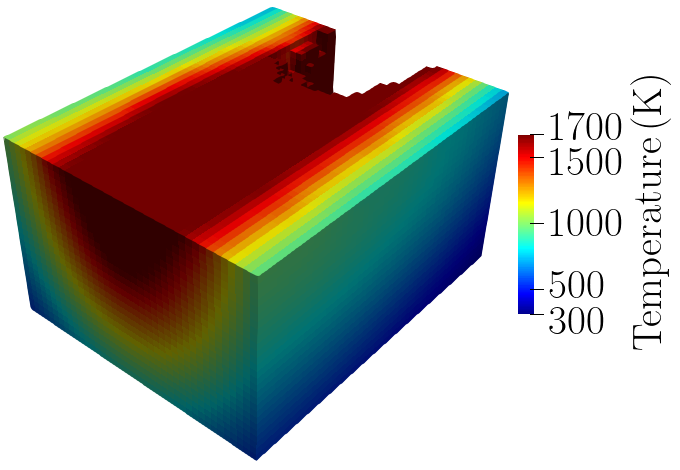}}
\newline
% Time 99 us
\subfloat[(d)]{\includegraphics[height=0.3\textwidth]{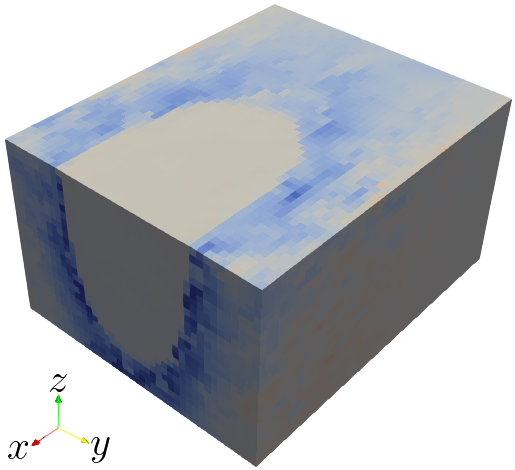}}
% Time 226 us
\subfloat[(e)]{\includegraphics[height=0.3\textwidth]{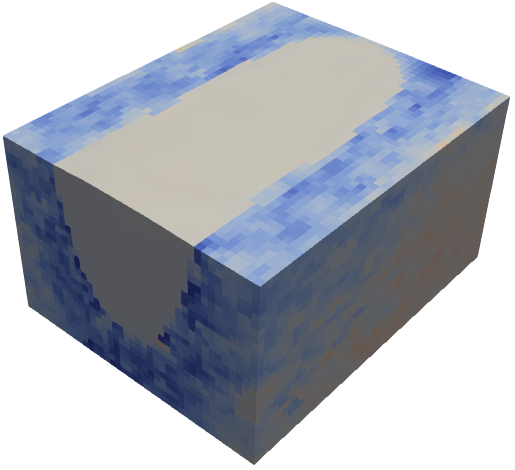}}
% Time 451 us
\subfloat[(f)]{\includegraphics[height=0.3\textwidth]{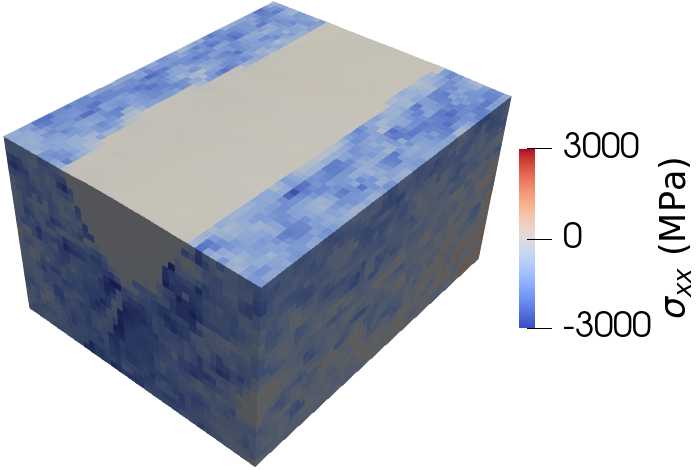}}
\caption{\label{fig:temperature} Snapshots of the temperature field using the element elimination and reactivation method at (a) $t = 99$ $\mu$s, (b) $t = 226$ $\mu$s, (c) $t = 451$ $\mu$s; (d)-(f) corresponding stress component $\boldsymbol{\sigma}_{xx}$ using the stiffness degradation method.}
% Temperature field:
% /home/nicolo/projects/c_pfor_am_Test/SpreadAndCFD/Residualstress4um/Case2DegradEigenBFI3_out/Case2DegradEigenBFI3_out.e
% stress field:
%/home/nicolo/projects/c_pfor_am_Test/SpreadAndCFD/Residualstress4um/ElementAddDelete/Case2StiffDegrad_out.e
%/home/nicolo/projects/c_pfor_am_Test/SpreadAndCFD/Residualstress4um/ElementAddDelete/Case2StiffDegradSigmaXX/SigmaXXtime451.png
%
\end{figure}

Crystal plasticity simulations of selective laser melting are carried out using the representative volume in Figure \ref{fig:grains}. A single scan track is modelled. The dimensions along the X, Y and Z axes are $200$ $\mu$m, $160$ $\mu$m and $108$ $\mu$m respectively. The FEM model represents a small region inside a larger sample. However, it is representative of the texture because of the large number of grains included. \\
The temperature field is obtained from thermal-fluid flow simulations
\cite{yan2017multi,wang2020evaporation,chen2020high}. The governing equations are the conservation of mass, momentum and energy. Most of the physical factors are incorporated, including the ray-tracing heat source model for laser, thermal conduction, surface radiation and convection, latent heat, evaporation, recoil pressure, viscosity, buoyancy force, surface tension and Marangoni effect. The details of the model and material parameters are reported in our previous works \cite{yan2017multi,wang2020evaporation}.  \\
The grain structure is obtained from a phase field model for grain growth \cite{YangMin2021}, where heterogeneous nucleation \cite{simmons2000phase} and the initial grain structures in powder particles and substrate are incorporated, and the same temperature profile from the thermal-fluid flow simulation is used. The phenomena, including grain nucleation and growth, competitive growth, epitaxial growth from powder particles and substrate, and grain coarsening in heat affected zones (HAZs), are comprehensively considered and experimentally validated. The details of the model and parameters are reported in our previous work \cite{YangMin2021}. \\
A total of 4522 grains are present with different orientations. The voxels obtained from the phase field model for grain growth are mapped directly into the FEM mesh, therefore the shape of the grains is reproduced accurately. The Euler angles $\varphi_1$, $\Phi$ and $\varphi_2$ are assigned based on electron backscatter diffraction (EBSD) experiments provided by Dr. Yin Zhang and published elsewhere \cite{chen2019microscale} to reproduce the realistic texture of AM 316 stainless steel. This is important because the grain orientation determines both the macroscopic and microscopic mechanical properties of the sample. On the other hand, the 3D shape of the grains cannot be fully extracted from EBSD measurements, therefore simulation results from the grain growth model are adopted. Figure \ref{fig:grains} shows the value of $\varphi_1$. The average grain size is approximately $16$ $\mu$m. However, grains in the centre of the laser track tend to be elongated along the $z$ axis, as shown in Figure \ref{fig:grains}. \\
The temperature field obtained from thermal fluid-flow simulations is projected on the mesh of the FEM simulations. Melting in the crystal plasticity simulations is described either by the stiffness degradation model described in section \ref{sec:residualstiffmethod} or by the element elimination and reactivation method described in section \ref{sec:elemelimreactimpl}. The temperature fields at different time steps are shown in Figure \ref{fig:temperature} (a)-(c), in which the eliminated elements are hidden from view. A structured mesh made of first order hexahedral elements is used. Simulations are carried out with the element size being $4$ $\mu$m or $8$ $\mu$m, respectively. The numbers of elements are 54000 and 7000 respectively. \\
The boundary conditions used in the simulations represent the constraints that are applied by the substrate and by the surrounding material. The displacement components are zero on the surfaces $x = 0$, $x = 200$ $\mu$m, $y = 0$ and $y = 160$ $\mu$m. The vertical displacement $\mathbf{u}_z$ is zero on the bottom surface, $z = 0$, to avoid translation along the $z$ axis. The displacement is free on the upper surface ($z = 108$ $\mu$m), where the laser beam is present.

\subsection{Simulation convergence}
\label{sec:simulationconvergence}

The convergence speed of the simulations using the stiffness degradation method depends mainly on the residual stiffness parameter $q_r$. A value smaller than 0.001 significantly increases the simulation time, but there is very small difference in the plastic deformation between the simulations with $q_r=0.01$ and $q_r=0.001$, as will be shown in the following. %\ref{sec:simulationresults} \hl{and Fig.} \ref{fig:plasticstraincentre}
Simulations with both $q_r=0.01$ and $q_r=0.001$ are carried out. \\
If the element elimination method is used, the time step must be decreased immediately before and after the temperature steps at which element elimination takes place, as described in section \ref{sec:cpmodel}. In this way, good convergence can be obtained. Specifically, the time step is reduced to $0.1$ $\mu$s in an interval of $2$ $\mu$s around each temperature step, i.e. every $45$ $\mu$s. For instance, considering the first temperature step, the time step is reduced to $0.1$ $\mu$s at $t = 44$ $\mu$s and it is increased again to $1$ $\mu$s after $t = 46$ $\mu$s. 
This time step reduction applies only during the laser scan, while time step is kept constant at $1$ $\mu$s during the subsequent cooling stage, which is the second part of the simulation, carried out after the laser beam has exited the geometry. By contrast, in the stiffness degradation method, the value of the time step is always $1$ $\mu$s. 
Since the eliminated elements do not contribute to the residual or Jacobian calculation of the solver, the computation of each time step is slightly faster if the element elimination method is used. However, a reduction of the time step size before and after the element elimination steps is necessary in that case. Therefore, the simulations carried out with the element elimination method are not necessarily faster, unless the number of eliminated elements is a significant fraction of the total number of elements. This is shown by the computational costs reported in table \ref{tab:simulationtime}. \\
\begin{figure}[!htb]
\centering
% Time 771 us
\subfloat[(a)]{\includegraphics[height=0.38\textwidth]{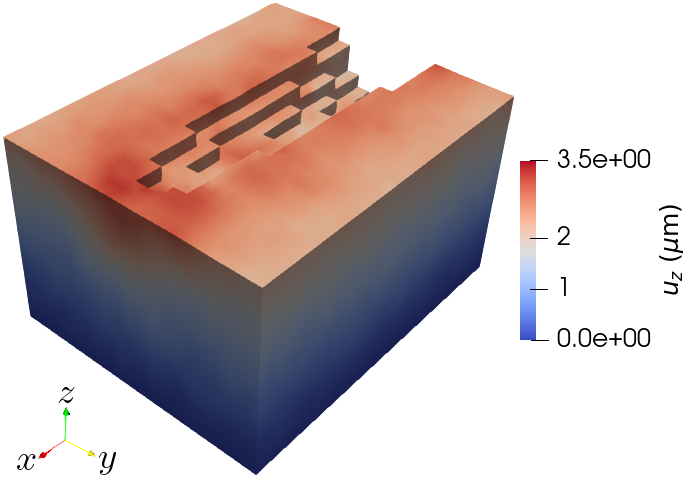}}
\subfloat[(b)]{\includegraphics[height=0.28\textwidth]{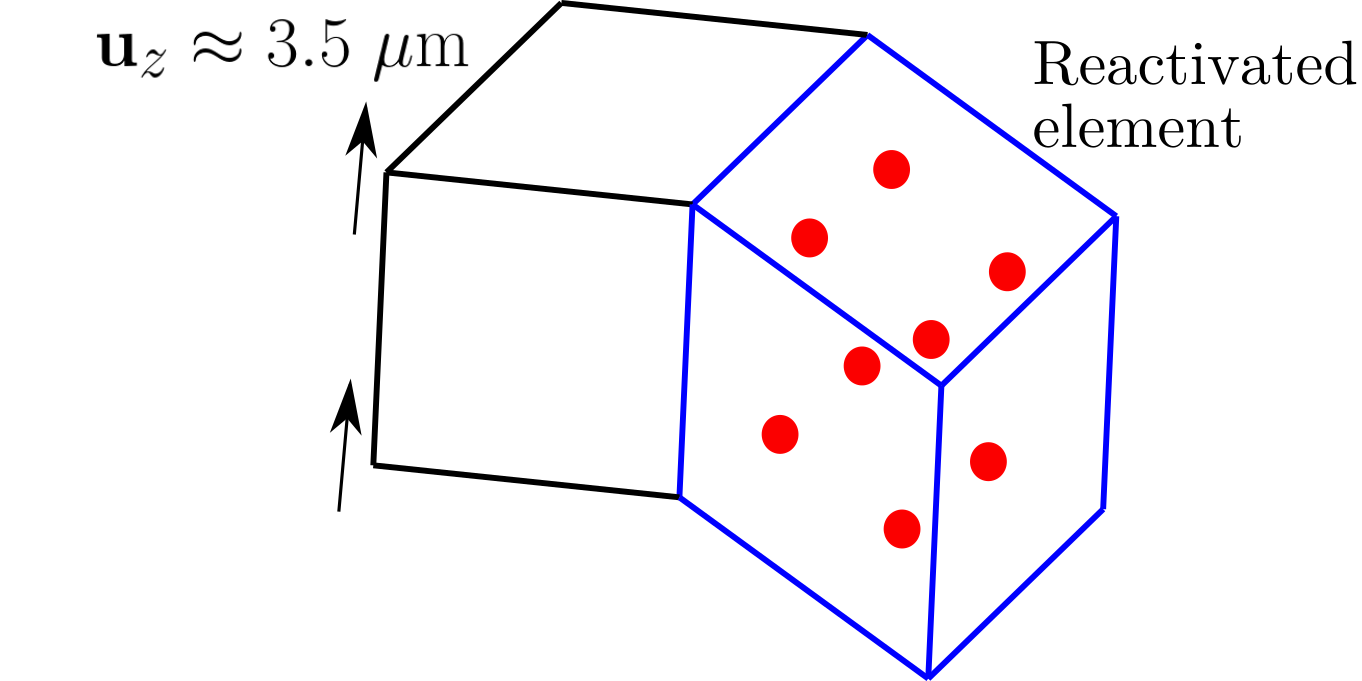}}
\newline
% Time 765 us
\subfloat[(c)]{\includegraphics[height=0.38\textwidth]{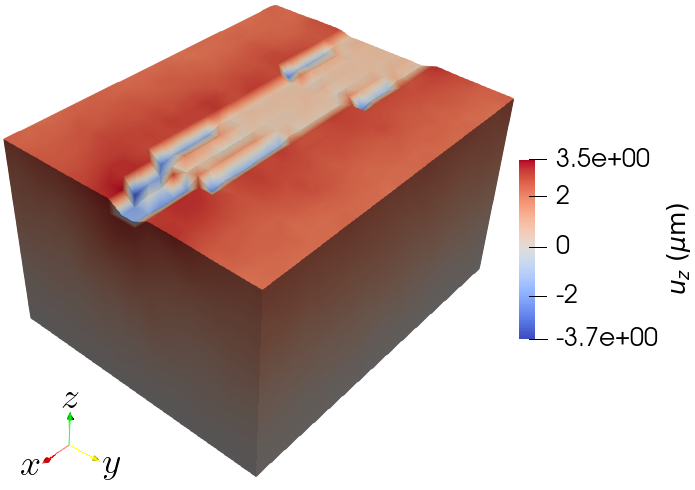}}
% Time 771 us
\subfloat[(d)]{\includegraphics[height=0.38\textwidth]{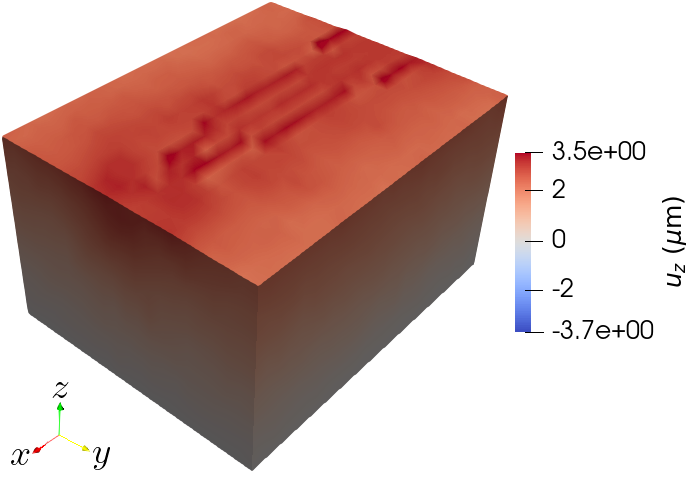}}
\caption{\label{fig:convergence} Vertical displacement $\mathbf{u}_z$ at $t = 771$ $\mu$s: (a) just before element reactivation, (c)-(d) just after element reactivation in case the reinitialisation $\mathbf{u}_z = 0$ or $\mathbf{u}_z = 3$ $\mu$m is used respectively. (b) Element distortion due to the reinitialisation $\mathbf{u}_z = 0$.}
% /home/nicolo/projects/c_pfor_am_Test/SpreadAndCFD/Residualstress8um/Case2DegradEigen_out/UzTime765_2.png
% /home/nicolo/projects/c_pfor_am_Test/ElementAddDelete/Slides/Images/ElementMemory3D.png
% /home/nicolo/projects/c_pfor_am_Test/SpreadAndCFD/Residualstress8um/Case2DegradEigenBFI3_out/UzTime771ElemDel.png
%
\label{fig:convergenceuz}
\end{figure}
\begin{table}[!htb]
\normalsize
\caption{Computational costs of the simulations.}
\label{tab:simulationtime}
\centering
\begin{tabular}{|c|c|c|c|} 
 \hline
 Mesh size & CPUs & Element elimination & Stiffness degradation \\ 
 \hline
 4 $\mu$m & 16 & 99 hours & 79 hours \\
 \hline 
 8 $\mu$m & 12 & 9.5 hours & 9.5 hours \\ \hline
\end{tabular}
\end{table}
As stated before, the element elimination steps require a reduction of the time step, however, they do not cause further convergence problems. The time steps at which element reactivation takes place leads to the following problem. Because of thermal expansion, there is a significant displacement along the $z$ axis on the top surface, as shown in Figure \ref{fig:convergenceuz}(a). Normally, reactivated elements have zero displacement because the displacement variable is reinitialised. In the MOOSE framework, the displacement in the reinitialised nodes is assigned in such a way as to get an average zero displacement at the eight integration points, as shown in Figure \ref{fig:convergenceuz}(b). Therefore, when an element is reactivated with zero displacement, the excessive distortion leads to divergence, as shown in Figure \ref{fig:convergenceuz}(c). \\
Several attempts have been carried out in which the initial conditions on the displacement after element reactivation are changed as a function of the coordinates. Surprisingly, a simple solution is to reinitialise $\mathbf{u}_z = 3$ $\mu$m when elements are reactivated, while $\mathbf{u}_x$ and $\mathbf{u}_y$ are reinitialised to zero everywhere. This displacement value is closer to the average displacement of the molten pool region on the top surface in the simulation using the residual stiffness method, while it also leads to good convergence. Other reinitialisation values for $\mathbf{u}_z$, such as 1 $\mu$m, 2 $\mu$m, 5 $\mu$m and $6$ $\mu$m have been used, but they have lead to excessive element distortion just after reinitialisation. Only values between 3 $\mu$m and 4 $\mu$m remove the element distortion problem, as shown in Figure \ref{fig:convergenceuz}(d). This implementation is made possible by the MOOSE framework because the initial conditions can be set as a function of time. At $t = 0$, the displacement is initialised to zero, while for $t > 0$, the components $\mathbf{u}_x$, $\mathbf{u}_y$ are reinitialised to zero and $\mathbf{u}_z$ is reinitialised to $3$ $\mu$m everywhere in the geometry. This strategy can be applied to any additive manufacturing simulation and can be extended by setting the initial conditions as a more complex function of space and time.

\subsection{Comparison between the computational methods}
\label{sec:simulationresults}

\begin{figure}[!htb]
\centering
\subfloat[(a)]{\includegraphics[height=0.4\textwidth]{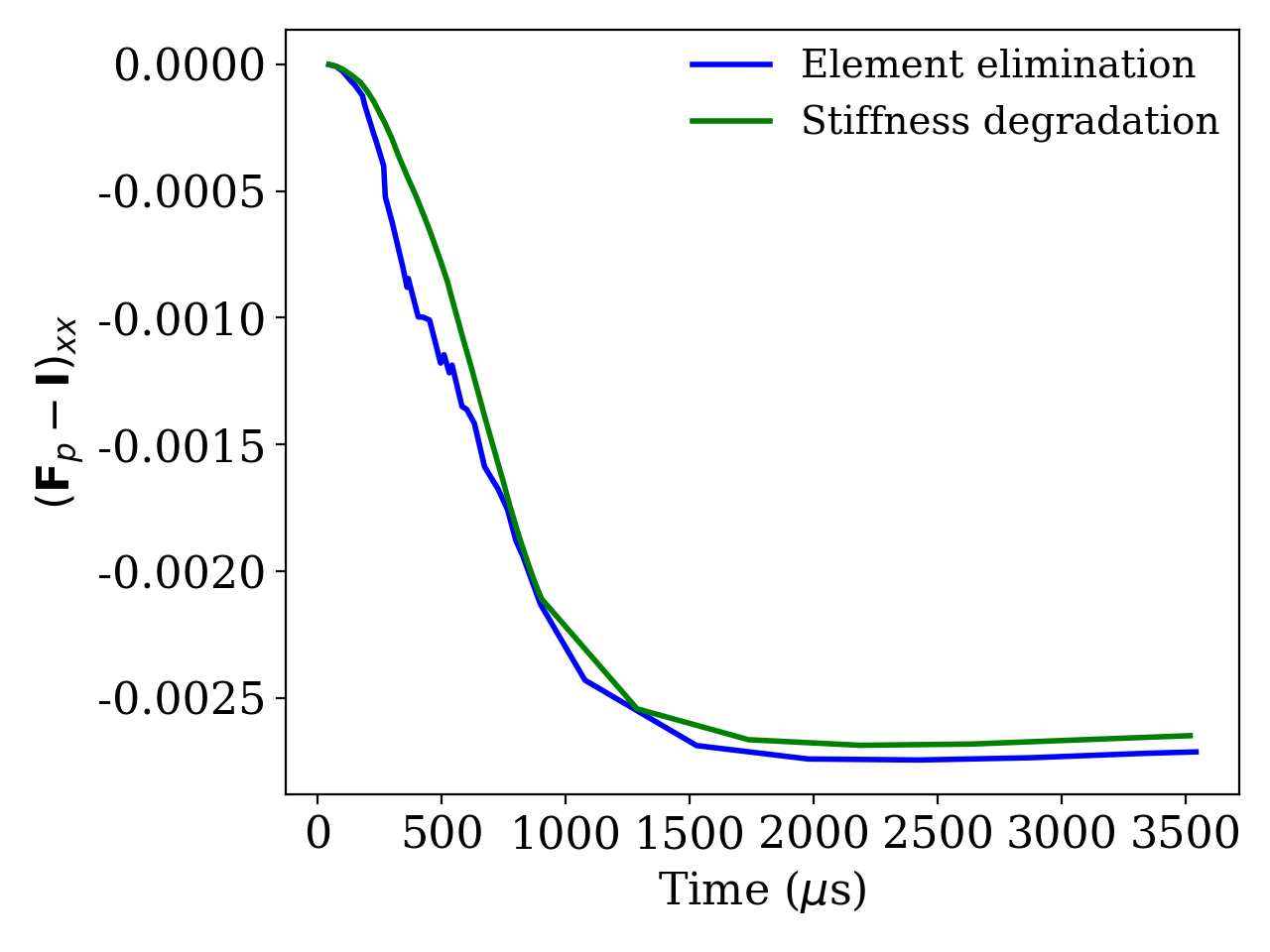}}
\subfloat[(b)]{\includegraphics[height=0.4\textwidth]{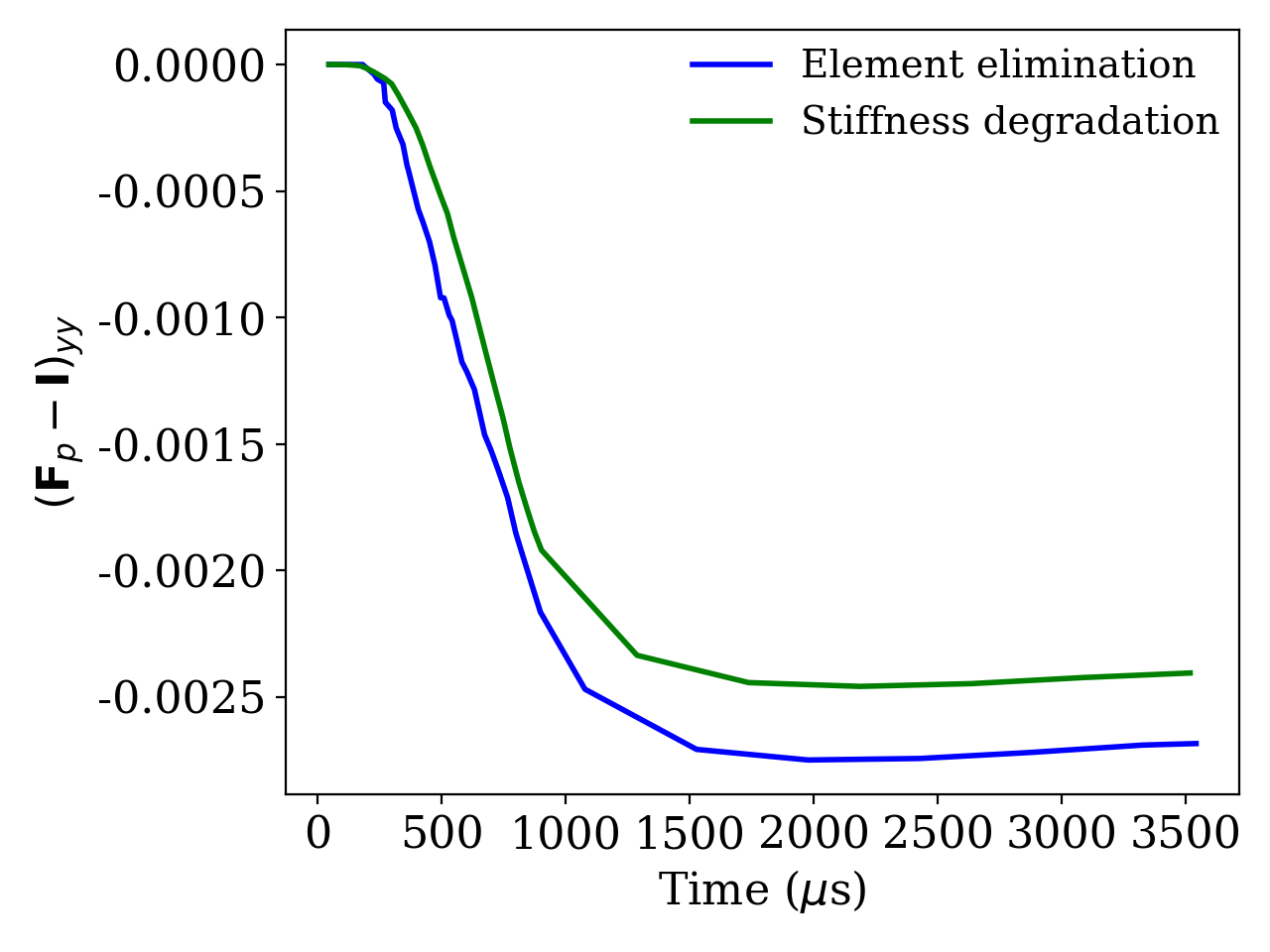}}
\newline
\subfloat[(c)]{\includegraphics[height=0.4\textwidth]{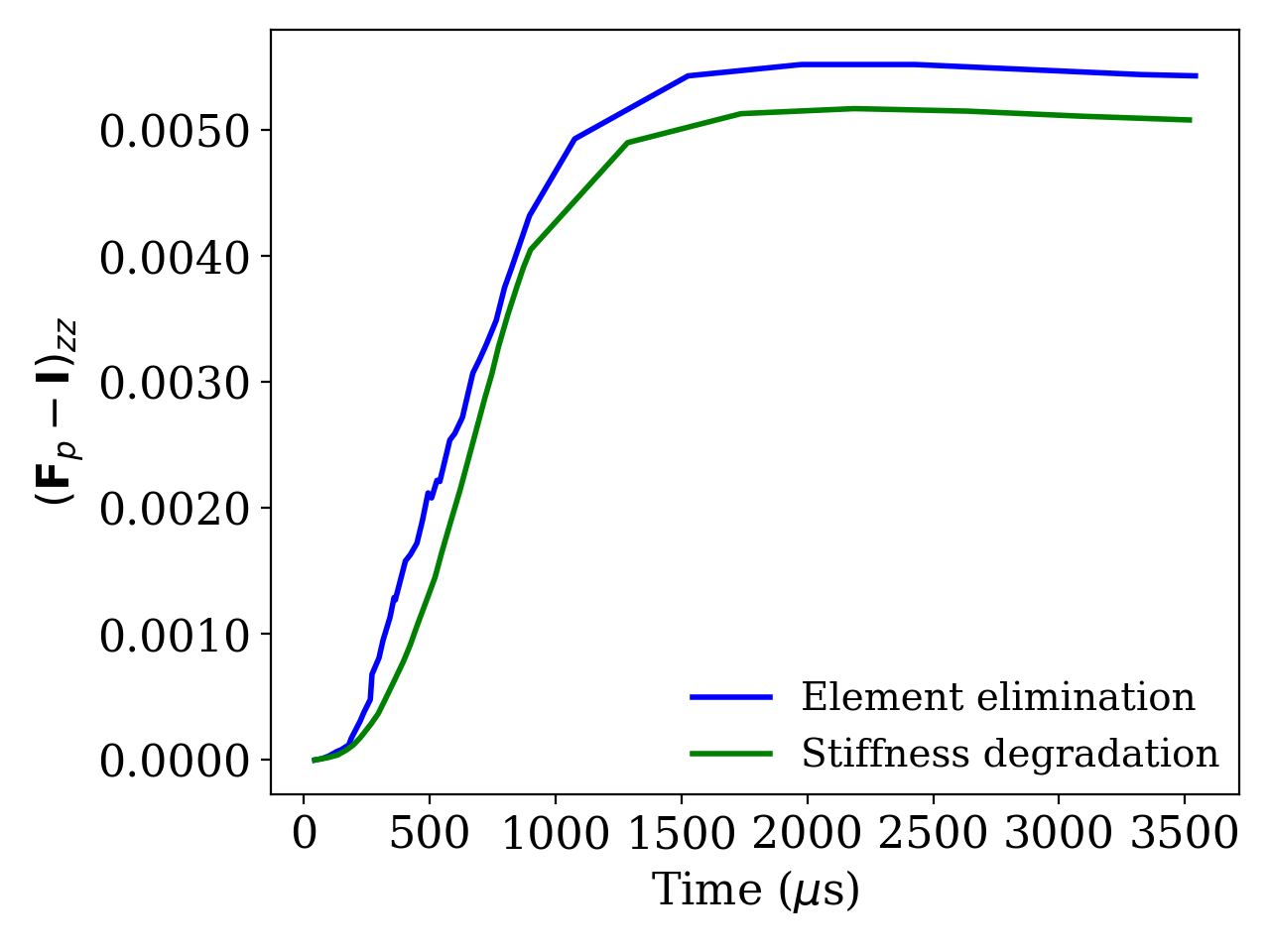}}
\subfloat[(d)]{\includegraphics[height=0.32\textwidth]{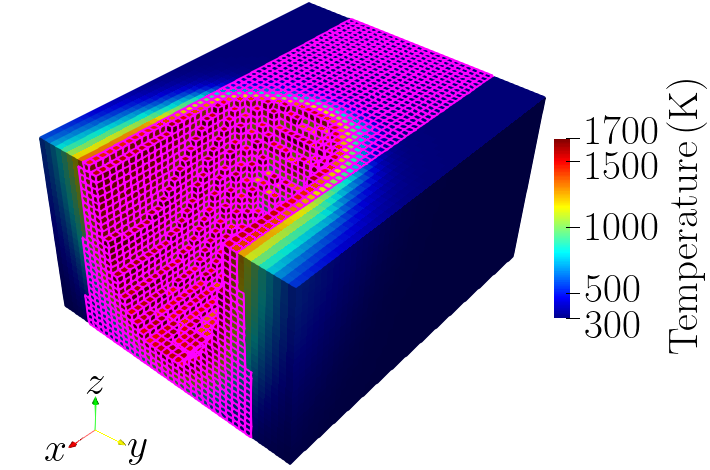}}
%/home/nicolo/projects/c_pfor_am_Test/SpreadAndCFD/Residualstress4um/Case2DegradEigenBFI3_out/Case2DegradEigenBFI3_out.e
%/home/nicolo/projects/c_pfor_am_Test/SpreadAndCFD/Residualstress4um/ElementAddDelete/Case2StiffDegrad_out.e
%/home/nicolo/projects/c_pfor_am_Test/SpreadAndCFD/Residualstress4um/Case2DegradEigenBFI3_out/SelectionCentre.png
\caption{\label{fig:plasticstraincentre} (a)-(c) Plastic strain components in the centre of the representative volume, averaged over the highlighted elements in (d).}
\end{figure}

\begin{figure}[!htb]
\centering
\subfloat[(a)]{\includegraphics[height=0.41\textwidth]{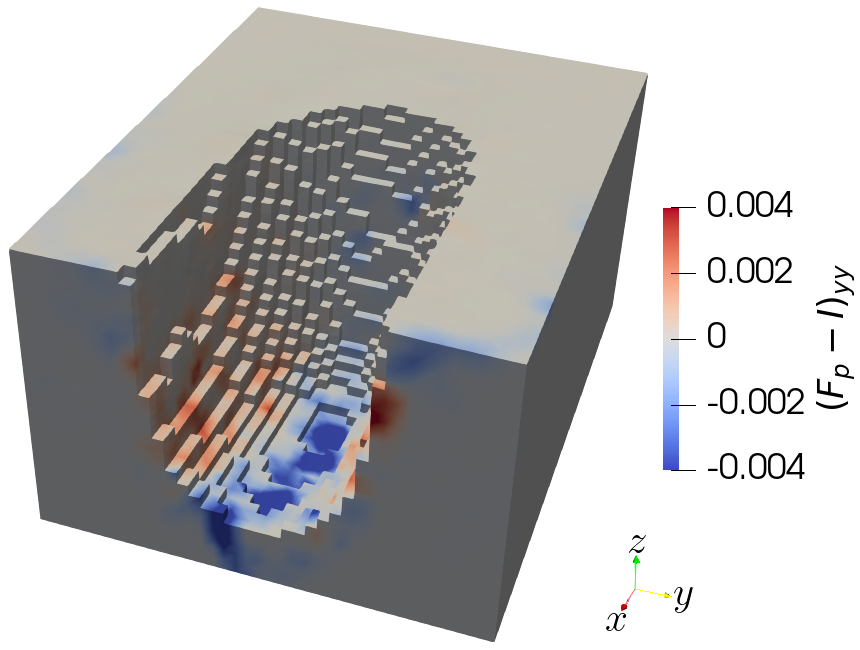}}
%\hspace{0.1cm}
\subfloat[(b)]{\includegraphics[height=0.41\textwidth]{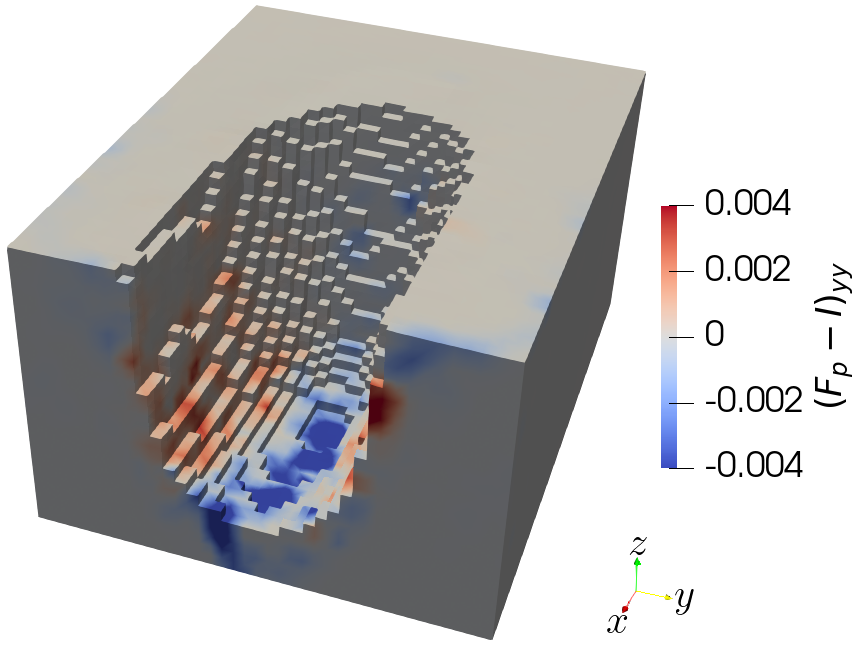}}
\caption{\label{fig:Fpqr} Component $\left ( \boldsymbol{F}_p \right )_{yy} - 1$ of the plastic deformation gradient at time $t = 179$ $\mu$s using the residual stiffness method with (a) $q_r=0.01$ and (b) $q_r=0.001$.}
% add path comparison 0.01 and 0.001
\end{figure}

The two computational methods presented in sections \ref{sec:residualstiffmethod} and \ref{sec:elemelimreactimpl} are compared. The plastic deformation is a very important quantity because it determines the residual stress after the laser scan. The diagonal components of $\boldsymbol{F}_p-\boldsymbol{I}$ are averaged over the central part of the representative volume, constituted of the highlighted elements in Figure \ref{fig:plasticstraincentre}(d). This procedure is carried out for simulations using the element elimination method and the stiffness degradation method. \\
Figures \ref{fig:plasticstraincentre}(a)-(c) show the comparison of simulation results by the two methods. During laser scan, the hot region undergoes thermal expansion and applies compression on the neighbouring regions. Therefore, plastic compression along the $x$ and $y$ axes takes place in the central region. Since plastic deformation is isochoric, the elements expand along the $z$ axis, as shown in Figure \ref{fig:plasticstraincentre}(c). \\
The plastic deformation is always larger using the element elimination and reactivation method. If the stiffness degradation method is used, the residual plastic deformation component along the $y$ axis, after laser scan and cooling, is underestimated by about 20\%. The difference is particularly important at time $400$-$500$ $\mu$s, when the laser just passes the representative volume but the melting pool is still deep. In this time interval, the stiffness degradation model underestimates the plastic deformation by a factor 2. \\
Overall, in the stiffness degradation method, the presence of low stiffness elements prevents the elements near the melting pool to accommodate part of the plastic deformation. Simulations with both $q_r=0.01$ and $q_r=0.001$ are carried out and the plastic deformation along the $y$ axis is shown in Figure \ref{fig:Fpqr} when the laser beam is approximately in the centre of the representative volume. The strong similarity between the plastic deformation in the two simulations confirms that the range of values chosen for the parameter $q_r$ does not affect the simulation results. The other components of the plastic deformation gradient are also very similar, therefore a value $q_r = 0.001$ is suitable for this kind of simulations.

\subsection{Grain orientation and plastic deformation}
\label{sec:grainoripdef}

\begin{figure}[!htb]
\centering
\subfloat[(a)]{\includegraphics[height=0.5\textwidth]{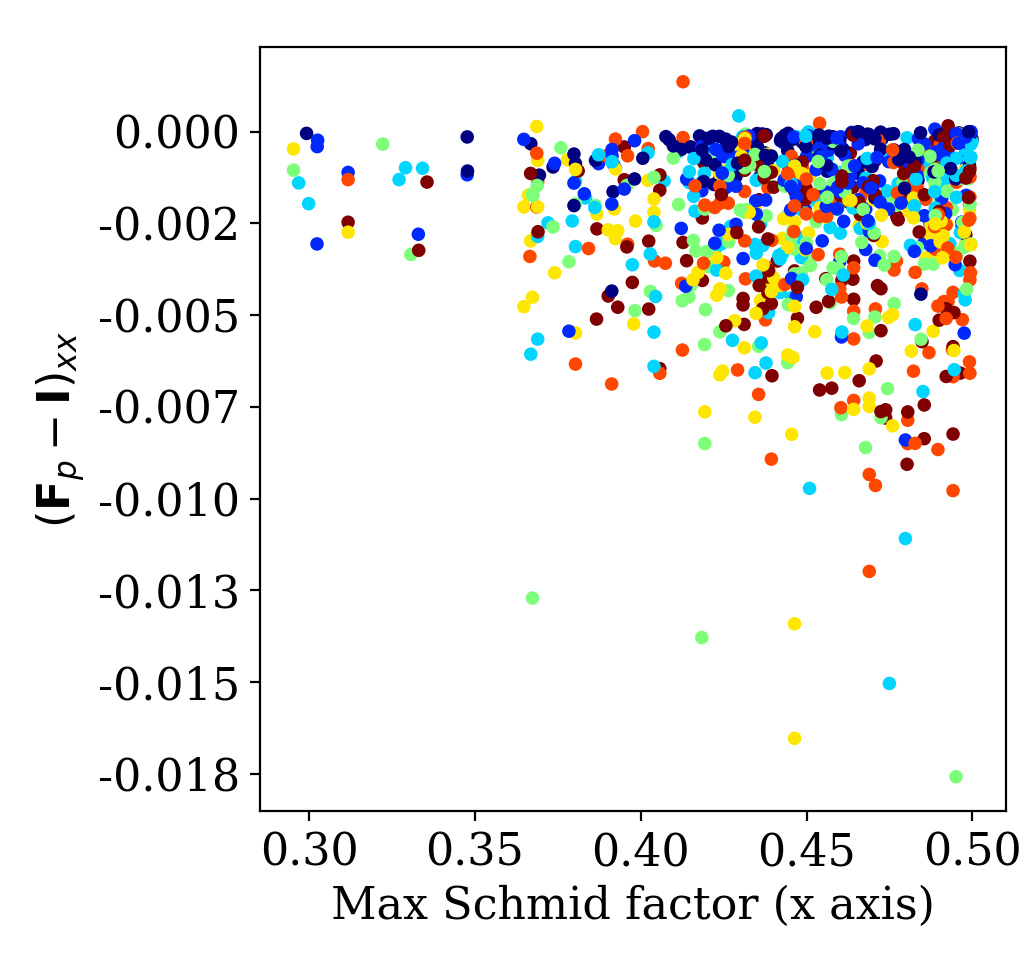}}
% /home/nicolo/projects/c_pfor_am_Test/SpreadAndCFD/Residualstress4um/Case2DegradEigenBFI3_out/FpxxSchmidFactorCoordsZLessP.png
\subfloat[(b)]{\includegraphics[height=0.5\textwidth]{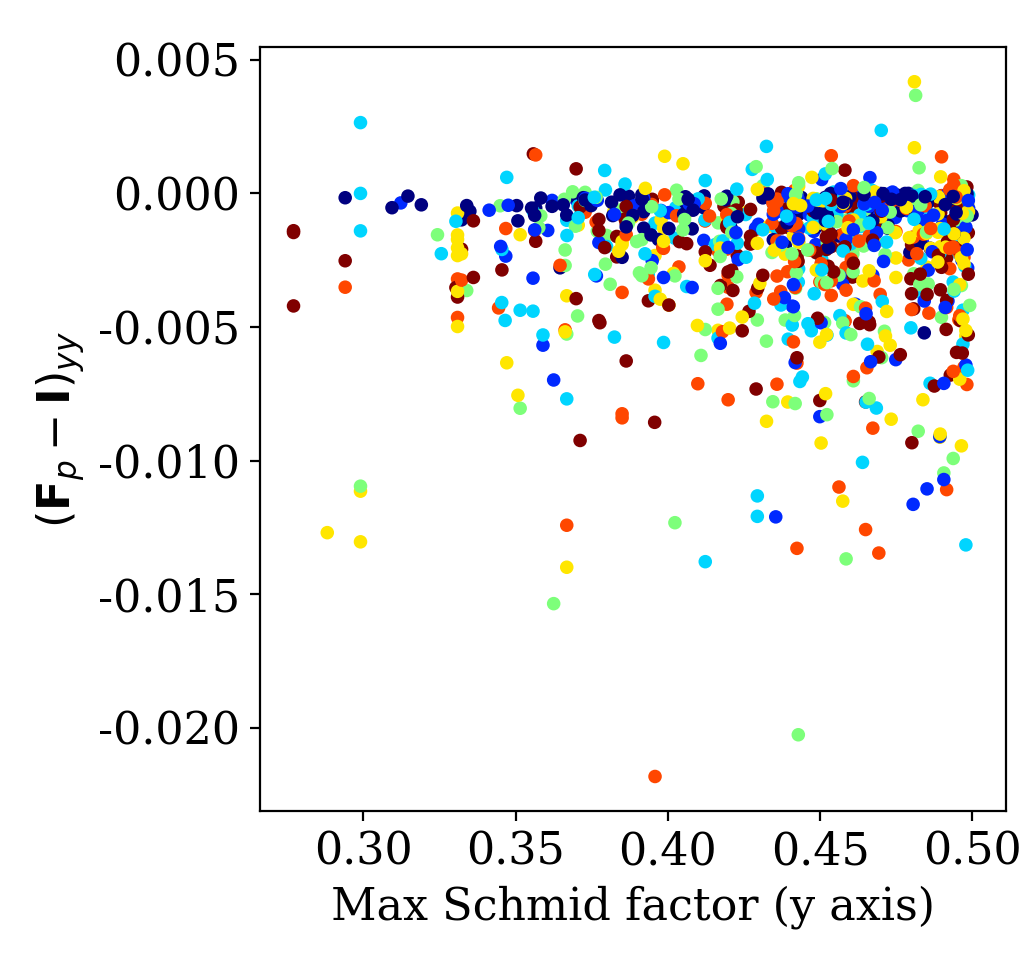}}
%/home/nicolo/projects/c_pfor_am_Test/SpreadAndCFD/Residualstress4um/Case2DegradEigenBFI3_out/FpyySchmidFactorCoordsZLessP.png
\newline
\subfloat[(c)]{\includegraphics[height=0.5\textwidth]{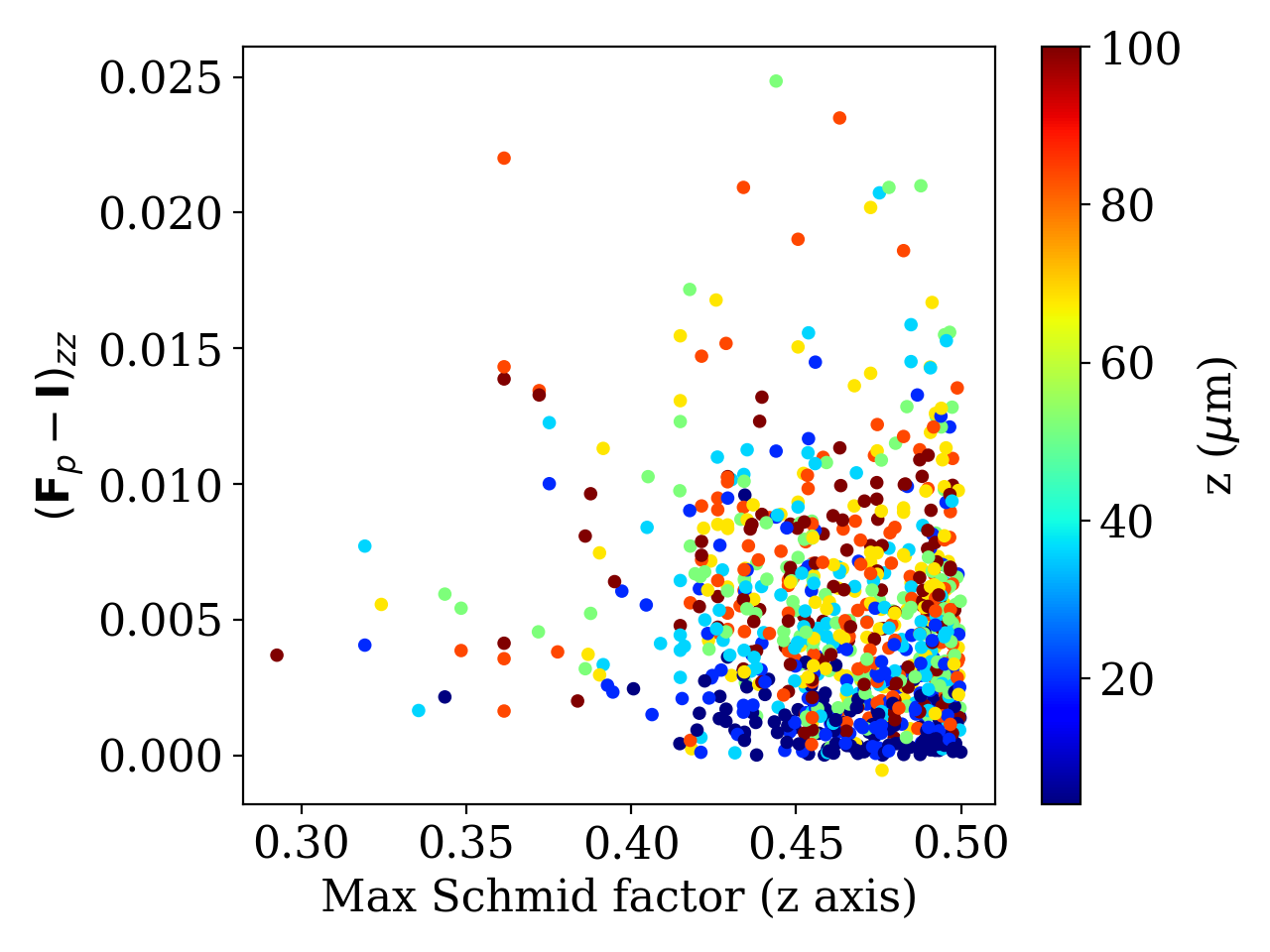}}
%/home/nicolo/projects/c_pfor_am_Test/SpreadAndCFD/Residualstress4um/Case2DegradEigenBFI3_out/FpzzSchmidFactorCoordsZLessP.png
\caption{\label{fig:schmidfactor} Correlation between maximum Schmid factor among the slip systems and plastic deformation after laser scan.}
%/home/nicolo/projects/c_pfor_am_Test/SpreadAndCFD/Residualstress4um/Case2DegradEigenBFI3_out/Case2DegradEigenBFI3_out.e
%/home/nicolo/projects/c_pfor_am_Test/SpreadAndCFD/Residualstress4um/Case2DegradEigenBFI3_out/CorrelationSchmidCoordsLessPoints.py
\end{figure}

The element elimination and reactivation method will be used in the following simulation results to understand the correlation between plastic deformation, residual stress and grain orientation. Firstly, the correlation between residual plastic deformation (after the laser scan and cooling) and grain orientation is analysed. The scatter plots in Figure \ref{fig:schmidfactor}(a)-(c) are obtained as follows: the plastic deformation components are averaged over each element in the representative volume and one element is selected every 16 $\mu$m and represented as a point in the scatter plots. The distribution of these elements is uniform in the representative volume and this algorithm allows to select a representative set of elements that do not fill completely the scatter plot. Given the Euler angles in each selected element, the maximum Schmid factor among the slip systems is calculated for load along a particular direction ($x$, $y$ or $z$ axes respectively) \cite{Grilli2016thesis}. This is motivated by the fact that, despite of the complex load undergone by different regions, the stress component along one direction should in principle induce plastic deformation along that particular direction. The colour of the points in Figure \ref{fig:schmidfactor}(a)-(c) corresponds to the position of the element along the $z$ axis. \\
\begin{figure}[!htb]
\centering
\subfloat[(a)]{\includegraphics[height=0.42\textwidth]{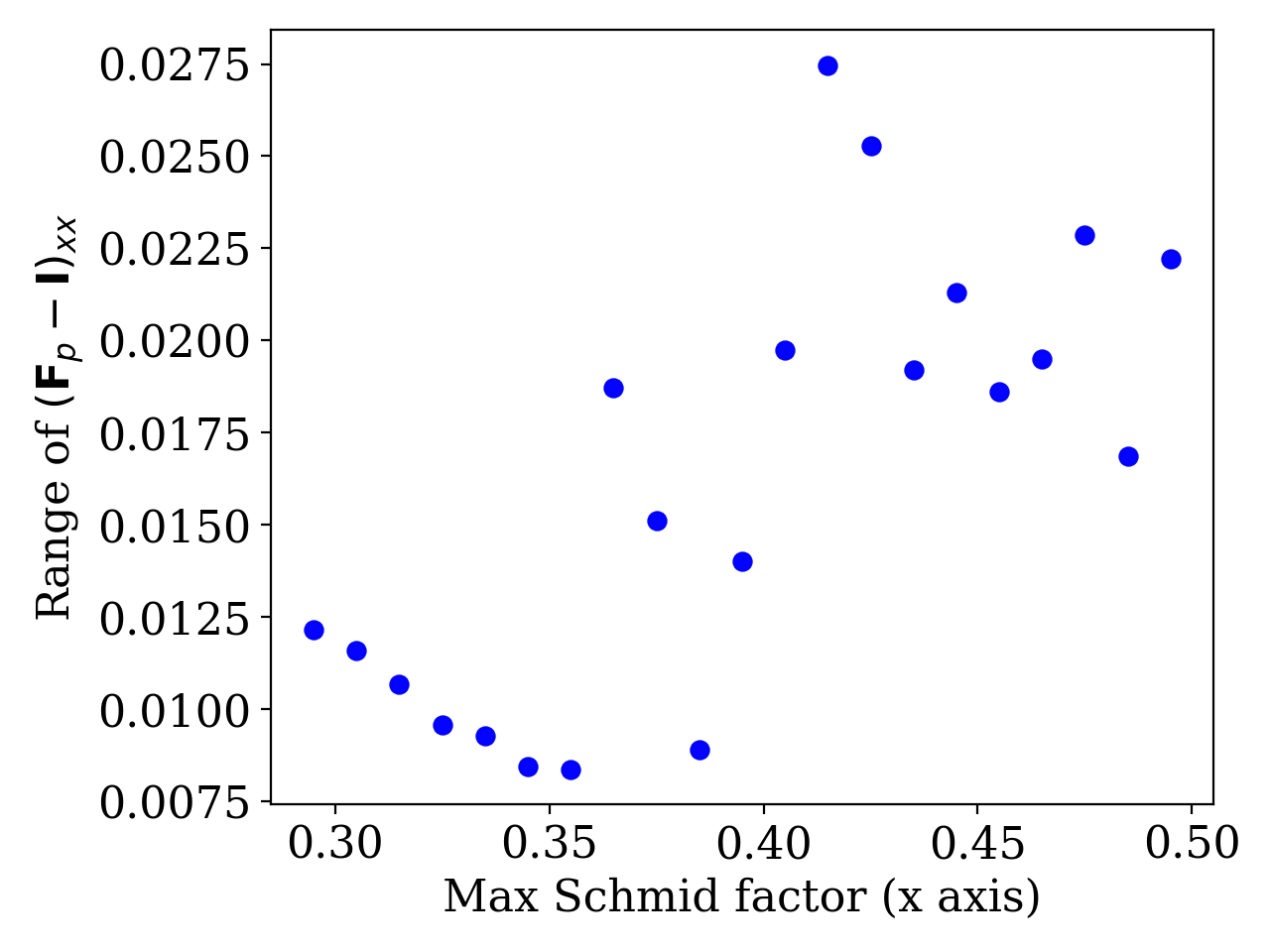}}
%/home/nicolo/projects/c_pfor_am_Test/SpreadAndCFD/Residualstress4um/Case2DegradEigenBFI3_out/FpxxRangeSchmidFactor.png
\subfloat[(b)]{\includegraphics[height=0.42\textwidth]{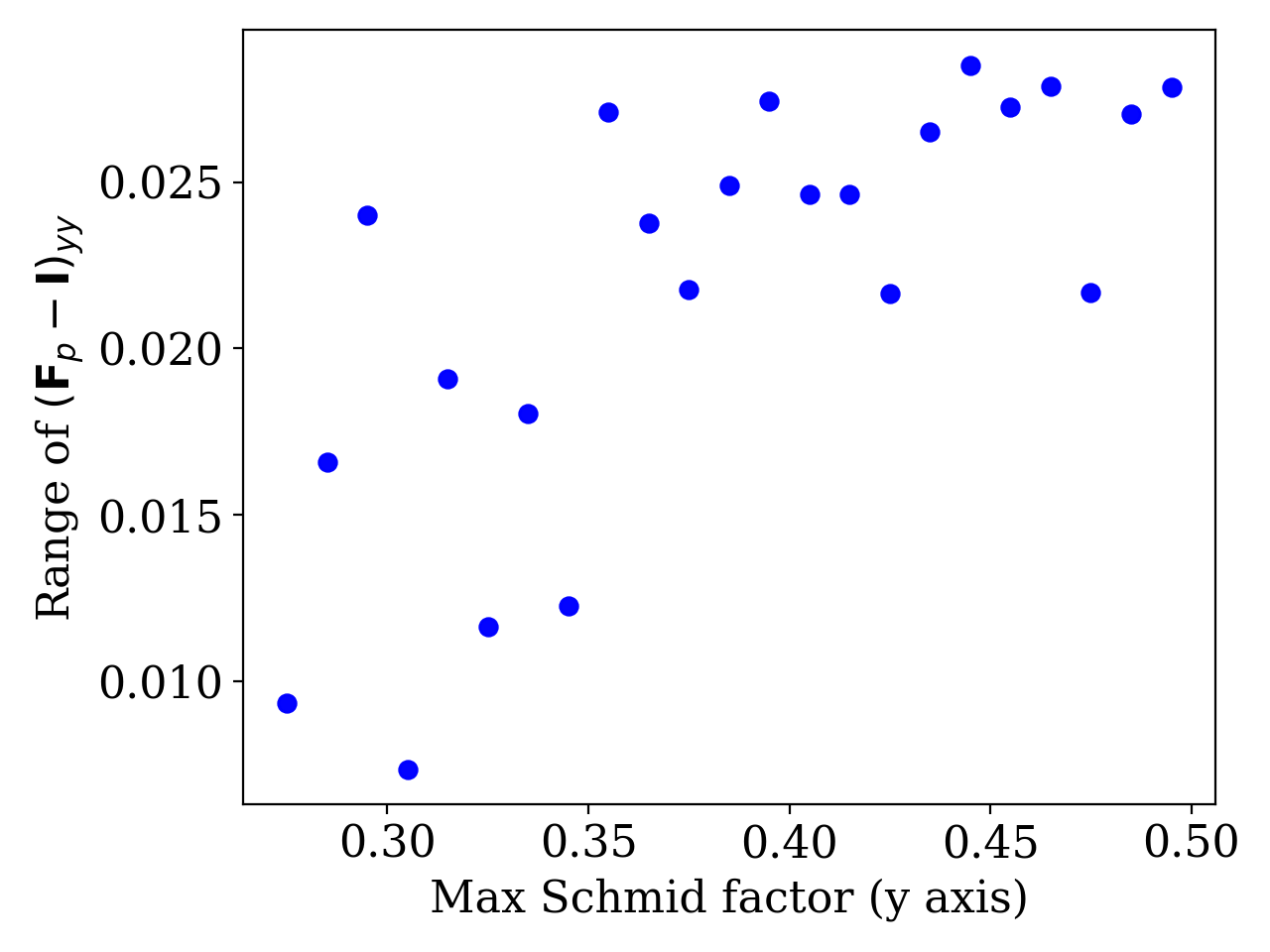}}
%/home/nicolo/projects/c_pfor_am_Test/SpreadAndCFD/Residualstress4um/Case2DegradEigenBFI3_out/FpyyRangeSchmidFactor.png
\newline
\subfloat[(c)]{\includegraphics[height=0.42\textwidth]{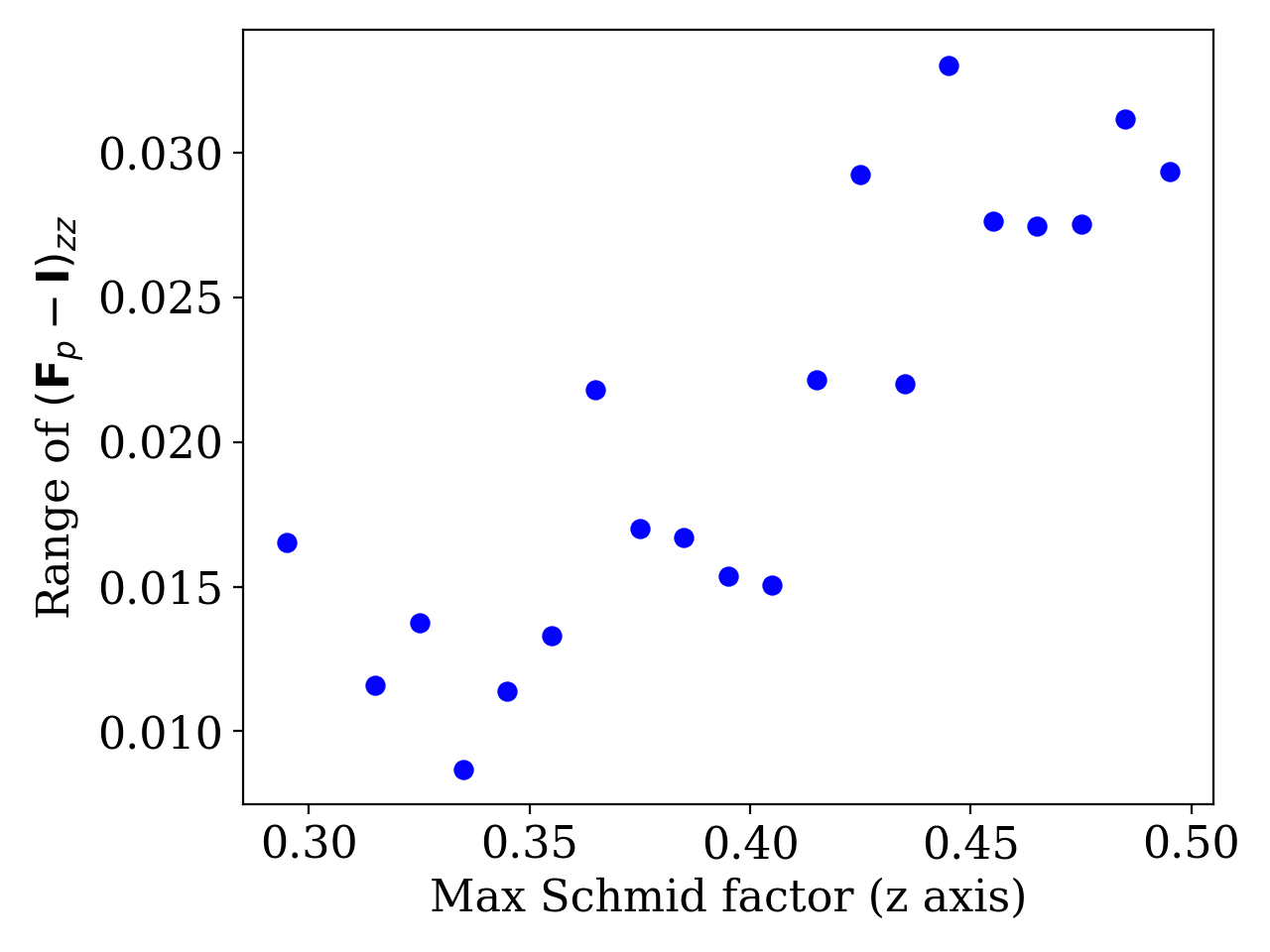}}
%/home/nicolo/projects/c_pfor_am_Test/SpreadAndCFD/Residualstress4um/Case2DegradEigenBFI3_out/FpzzRangeSchmidFactor.png
\caption{\label{fig:rangeschmidfactor} Correlation between maximum Schmid factor among the slip systems and the range of plastic deformation after laser scan.}
\end{figure}
As shown in Figure \ref{fig:schmidfactor}, the maximum plastic deformations occur on some points that are at an intermediate depth (about 40 $\mu m$ to 80 $\mu m$). This is related to the solidification and thermal stress around the molten pool, as also revealed in Figure \ref{fig:FpMeltPool}. Moreover, it is clear that a higher number of grains with larger maximum Schmid factors along a specific direction ($x$, $y$ or $z$) tends to have larger plastic deformation along that direction. However, there are grains with large maximum Schmid factors that show little plastic deformation. Many of this kind of points are dark blue in Figure \ref{fig:schmidfactor}(a)-(c), which means they are near the substrate. Because the substrate is fixed during simulation, the plastic deformation near the substrate is limited. Nevertheless, there are still some points at intermediate depth that do not have much plastic deformation, this is quite surprising and reflects the complexity of the thermal load in different regions. This behaviour will be clarified in the following, when the plastic deformation during laser scan is shown near the melting pool. There are also some grains with larger Schmid factor that show little plastic deformation and are located near the top stress-free surface (dark red points), as shown in Figure \ref{fig:schmidfactor}(a)-(c). \\
To investigate the directionality of the plastic deformation, the range of the plastic deformation components is investigated, which shows a greater correlation with the maximum Schmid factor \cite{IRASTORZALANDA2017157}, as shown in Figure \ref{fig:rangeschmidfactor}(a)-(c). The range is calculated as the difference between maximum and minimum values of the plastic strain components for a given maximum Schmid factor. 
%Although the stress state is complicated in the domain, there is still laser-induced prone feature of the stress distribution, thus the correlation between the plastic deformation components and the loading directions on the grains is investigated.
This correlation does not depend strongly on the depth $z$. 
The majority of the grains undergo plastic compression along the $x$ and $y$ axes, while plastic tension is observed along the $z$ axis. This is consistent with the average plastic strain in Figure \ref{fig:plasticstraincentre}.

\subsection{Plastic deformation and residual stress}
\label{sec:pdefresstress}

\begin{figure}[!htb]
\centering
\subfloat[(a)]{\includegraphics[height=0.5\textwidth]{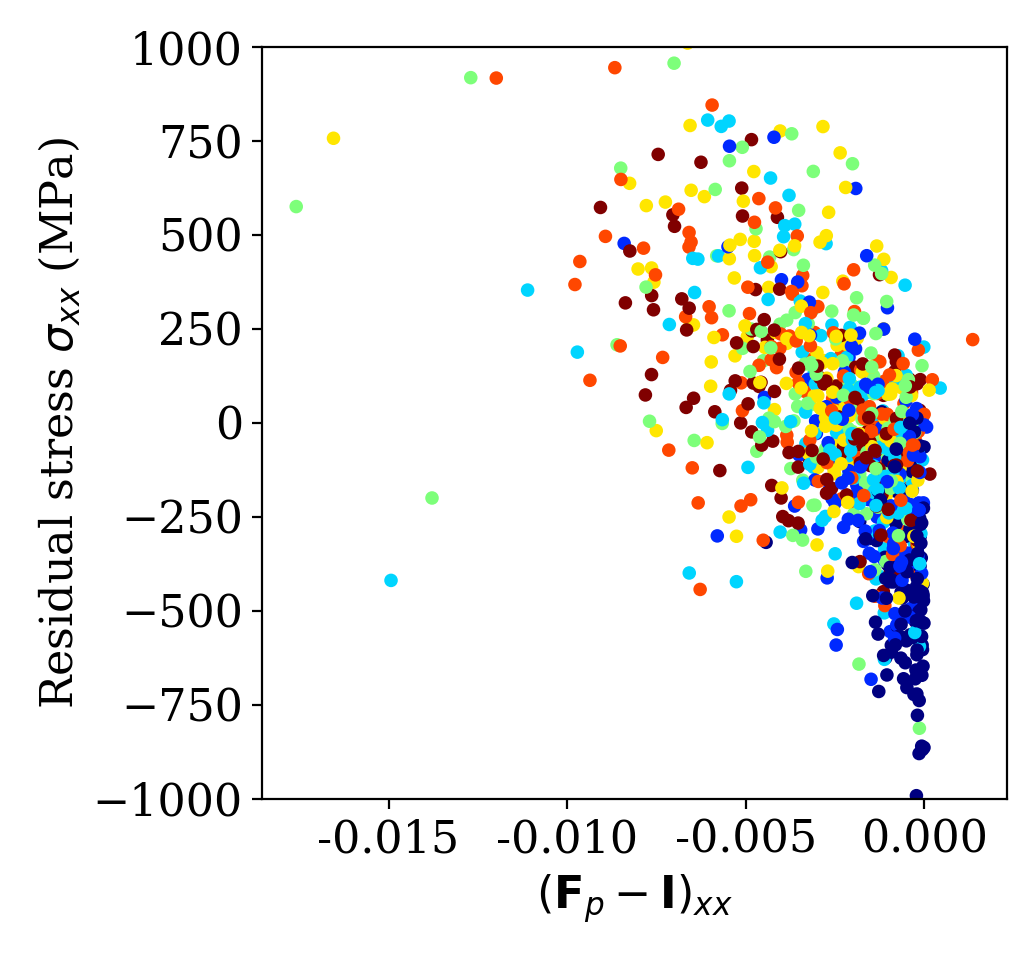}}
\subfloat[(b)]{\includegraphics[height=0.5\textwidth]{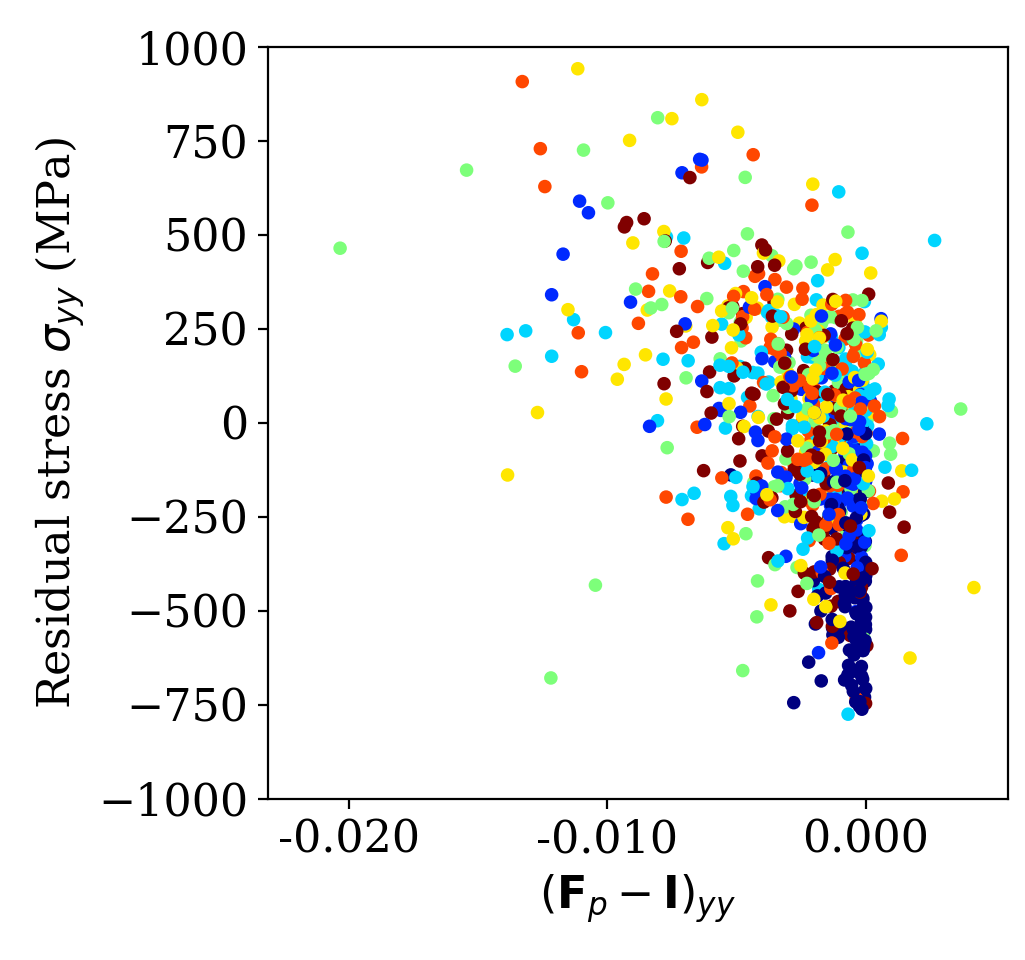}}
\newline
\subfloat[(c)]{\includegraphics[height=0.5\textwidth]{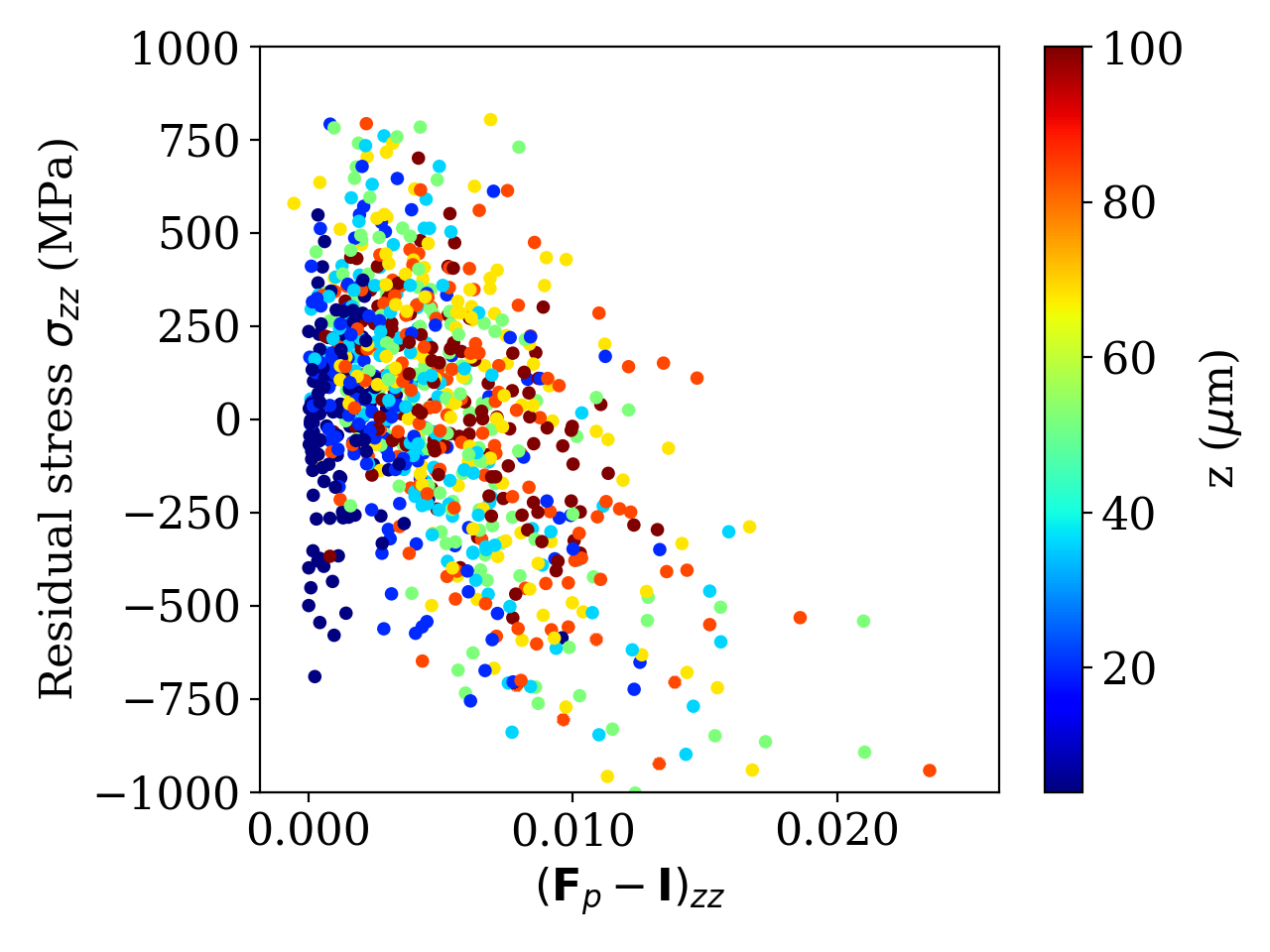}}
% /home/nicolo/projects/c_pfor_am_Test/SpreadAndCFD/Residualstress4um/Case2DegradEigenBFI3_out/Case2DegradEigenBFI3_out.e
%/home/nicolo/projects/c_pfor_am_Test/SpreadAndCFD/Residualstress4um/Case2Relaxation/Case2RelaxationDisp_out.e
%/home/nicolo/projects/c_pfor_am_Test/SpreadAndCFD/Residualstress4um/Case2Relaxation/CorrelationResStressCoordsLessPoints.py
%/home/nicolo/projects/c_pfor_am_Test/SpreadAndCFD/Residualstress4um/Case2Relaxation/ResStressXXCoordsZLessP.png
%/home/nicolo/projects/c_pfor_am_Test/SpreadAndCFD/Residualstress4um/Case2Relaxation/ResStressYYCoordsZLessP.png
%/home/nicolo/projects/c_pfor_am_Test/SpreadAndCFD/Residualstress4um/Case2Relaxation/ResStressZZCoordsZLessP.png
\caption{\label{fig:resstress} Correlation between the components of the residual stress and components of the plastic deformation gradient.}
\end{figure}

\begin{figure}[!htb]
\centering
\subfloat[(a)]{\includegraphics[height=0.4\textwidth]{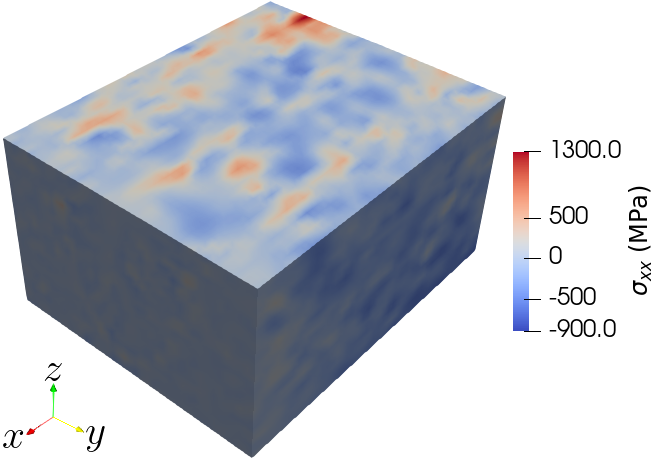}}
\subfloat[(b)]{\includegraphics[height=0.4\textwidth]{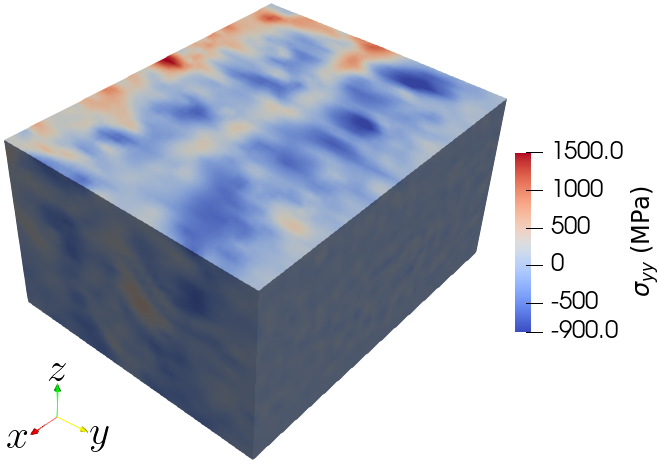}}
\newline
\subfloat[(c)]{\includegraphics[height=0.42\textwidth]{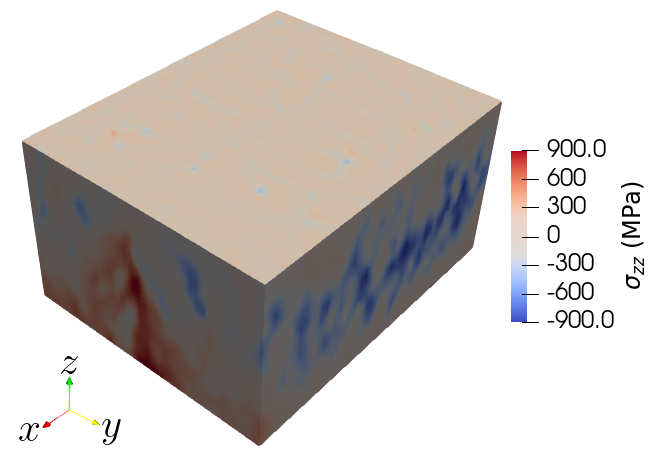}}
\caption{\label{fig:SigmaFpMeltPool} Residual stress components after the cooling stage of the simulation.}
\end{figure}

The residual stress, after laser scan and cooling, is due to the heterogeneity of the plastic strain in different grains. In order to maintain strain compatibility, the elastic deformation must compensate the difference in plastic deformation, therefore residual stresses are generated, as modelled by equations (\ref{eqn:elastoplasticdecomposition}) and (\ref{eqn:pkstress}). 
The correlation between residual stress and plastic deformation
can be visualised using the scatter plots in Figure \ref{fig:resstress}.
Each point represents an element in the representative volume, selected every 16 $\mu$m,
with its corresponding residual stress and plastic strain, averaged over the integration points.
The colour of each point represents the $z$ coordinate of the corresponding element. \\
It is certainly true that there is a correlation between residual stress and plastic strain.
Compressive plastic strain along the $x$ and $y$ axes corresponds to tensile residual stress components
along those directions, while tensile plastic strain along the $z$ axis corresponds to
compressive residual stress along $z$. However, few points show the opposite behaviour: for instance, a tensile residual stress $\boldsymbol{\sigma}_{yy}$ is associated with a small tensile plastic strain
along the $y$ axis in some regions. Overall, the tensile residual stress components $\boldsymbol{\sigma}_{xx}$ and $\boldsymbol{\sigma}_{yy}$ are similar and large, therefore these pre-tensions can induce fracture along both the $x$ and $y$ directions. \\
The plastic deformation depends also on the depth. In Figure \ref{fig:resstress}, it is evident that
the regions at the top of the representative volume, represented by dark red points,
are not the ones accommodating the largest plastic strain. The points with the largest
plastic deformation are the ones with depth in the interval 30 $\mu$m $< z < 80$ $\mu$m.
These are the points that are closer to the bottom of the melting pool during laser scan.
However, the points at the top surface can have residual stresses that are similar
to points that undergo larger plastic deformations. This shows that the influence of the plastic deformation in the substrate is very important; it will be discussed in more details in section \ref{sec:discussion}. It is important to note that some elements show values of the residual stress components that are larger than the critical resolved shear stress to induce plastic deformation. This is due to large volumetric stress in those elements. Moreover, some dark red points in Figure \ref{fig:resstress}(c) have residual stress $\sigma_{zz}$ that is different from zero. These points are picked just below the surface, in the coordinate interval 90 $\mu$m $< z < 100$ $\mu$m. Indeed the stress component $\sigma_{zz}$ on the free top surface is very close to zero, as shown in Figure \ref{fig:SigmaFpMeltPool}(c). \\
\begin{figure}[!htb]
\centering
\subfloat[(a)]{\includegraphics[height=0.41\textwidth]{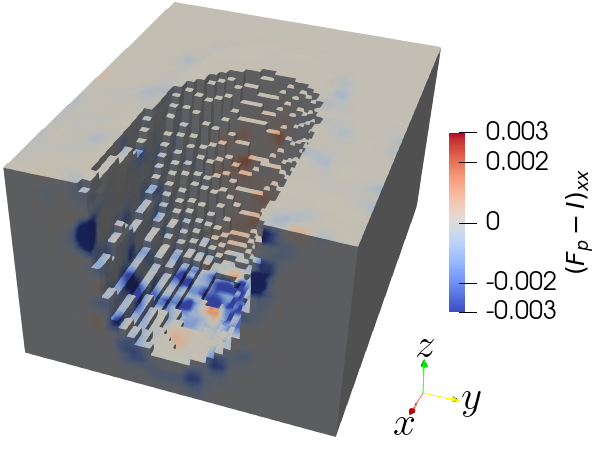}}
%\hspace{0.1cm}
\subfloat[(b)]{\includegraphics[height=0.41\textwidth]{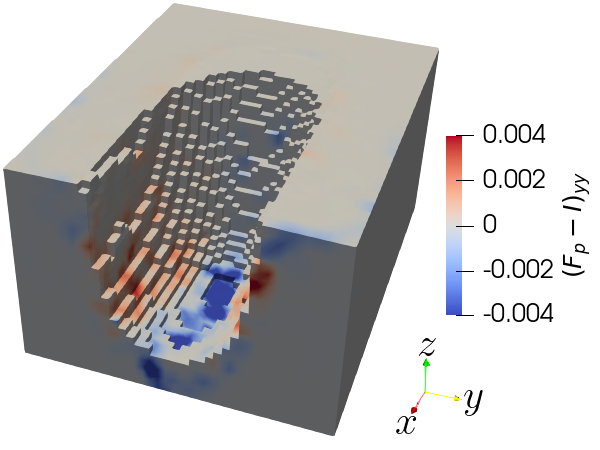}}
\newline
\subfloat[(c)]{\includegraphics[height=0.41\textwidth]{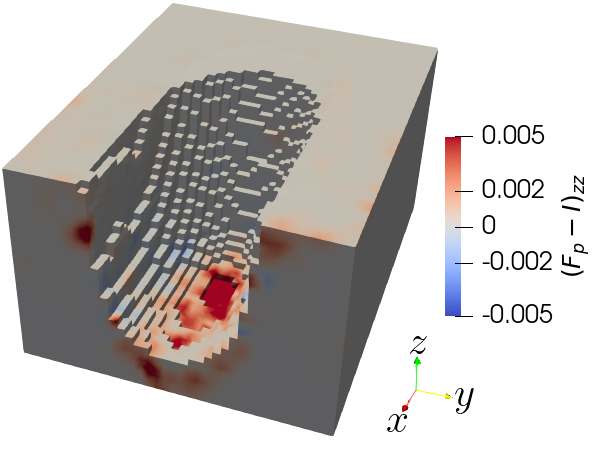}}
\caption{\label{fig:FpMeltPool} Components of the plastic deformation gradient $\boldsymbol{F}_p$ at the bottom of the melting pool at time $t = 179$ $\mu$s.}
%%/home/nicolo/projects/c_pfor_am_Test/SpreadAndCFD/Residualstress4um/Case2DegradEigenBFI3_out/FpView.pvcc
%%/home/nicolo/projects/c_pfor_am_Test/SpreadAndCFD/Residualstress4um/Case2DegradEigenBFI3_out/Case2DegradEigenBFI3_out.e
\end{figure}
% /home/nicolo/projects/c_pfor_am_Test/SpreadAndCFD/Residualstress4um/Case2Relaxation/Case2RelaxationDisp_out.e
% /home/nicolo/projects/c_pfor_am_Test/SpreadAndCFD/Residualstress4um/Case2Relaxation/StressZZ.pvcc
%/home/nicolo/projects/c_pfor_am_Test/SpreadAndCFD/Residualstress4um/Case2Relaxation/StressZZ.pvsm
In order to understand the plastic deformation in the representative volume induced by the laser scan, the components of $\boldsymbol{F}_p$ are shown in Figure \ref{fig:FpMeltPool}(a)-(c) when the laser beam is approximately in the centre of the representative volume. 
The thermal expansion ahead of the laser beam leads to plastic compression along the $x$ axis. This is visible on the two sides and at the bottom of the melting pool in Figure \ref{fig:FpMeltPool}(a), where blue areas are present. 
By contrast, the plastic deformation along the $y$ axis is compressive at the bottom of the melting pool but it is tensile on the two sides, as shown by the red regions in Figure \ref{fig:FpMeltPool}(b). 
Since plastic deformation is isochoric, the region at the bottom of the melting pool expands along the $z$ axis, as shown in Figure \ref{fig:FpMeltPool}(c). Because of the subsequent solidification, the melting pool becomes more shallow and the plastic deformation concentration shown in Figure \ref{fig:FpMeltPool}(a)-(c) spreads in the central region of the representative volume. This is the reason why the regions with the largest plastic deformation are concentrated at depth 30 $\mu$m $< z < 80$ $\mu$m, as shown in Figure \ref{fig:resstress}. 
At the top free surface, the plastic deformation is slightly smaller than in the centre of the representative volume, as shown in Figure \ref{fig:schmidfactor}(a)-(c). This is the reason why not all grains with larger Schmid factor at the top surface can accommodate large plastic deformations, as stated in section \ref{sec:grainoripdef}. \\
The diagonal components of the residual Cauchy stress tensor are shown in Figures \ref{fig:SigmaFpMeltPool}(a)-(c). $\boldsymbol{\sigma}_{xx}$ and $\boldsymbol{\sigma}_{yy}$ are heterogeneous at the top surface and in the centre of the representative volume, as shown in Figures \ref{fig:SigmaFpMeltPool}(a) and (b). This shows that they are strongly affected by the grain orientation and that the substrate is applying an important constraint along $x$ and $y$ on the top surface. In fact, even if the top surface is free, large values of $\boldsymbol{\sigma}_{xx}$ and $\boldsymbol{\sigma}_{yy}$ are present. The compressive stress $\boldsymbol{\sigma}_{zz}$ is concentrated in the substrate but it is not very large at the top surface because of the free boundary condition.

\section{Discussion}
\label{sec:discussion}

\begin{figure}[!htb]
\centering
\includegraphics[width=1.0\textwidth]{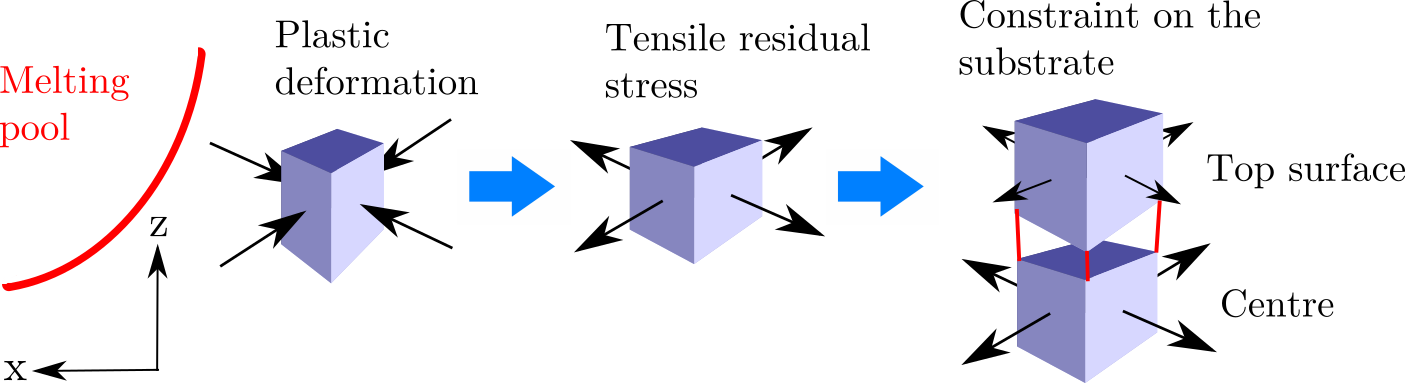}
\caption{\label{fig:constraintsubstrate} Mechanism for the build up of the residual stresses.}
% /home/nicolo/projects/c_pfor_am_Test/CFDGrainGrowth/Slides/Images/ConstraintSubstrate.svg
\end{figure}

The simulations suggest the following mechanism for the build up of the residual stresses, as shown in Figure \ref{fig:constraintsubstrate}. The thermal expansion around the molten pool induces plastic compression along the horizontal directions, $x$ and $y$, and plastic tension along the out-of-plane $z$ direction. This is particularly obvious at the bottom of the melting pool, as shown in Figure \ref{fig:FpMeltPool}. After the laser scan is over, tensile residual stresses build up along $x$ and $y$ directions to compensate the plastic compression. This effect is more relevant in the centre of the representative volume, however constraints are present between the regions close to the top surface and the centre. This induces residual stresses along the $x$ and $y$ directions also close to the top surface, as shown in Figure \ref{fig:SigmaFpMeltPool}. \\
The different grain orientations cause the heterogeneity of the residual stress. Grains with slip systems favourably oriented for plastic slip during loads along the $x$, $y$ and $z$ directions are more likely to develop plastic deformation along the corresponding directions. The simulations show that the maximum Schmid factor along the three axes is a suitable quantity to predict the corresponding plastic deformation and consequent residual stress, as shown in Figures \ref{fig:schmidfactor} and \ref{fig:resstress}. \\
The simulations show that, in the stiffness degradation method, the presence of regions with a small residual stiffness imposes a constraint on the regions near the melting pool, preventing more plastic deformation to take place. This is the origin of the underestimation of the plastic deformation and of the residual stress using the stiffness degradation method. By contrast, the element elimination and reactivation method predicts higher plastic deformation in selective laser melting simulations. This difference between the two methods must be taken into account in future studies. \\
Regions with larger tensile residual stresses along specific directions are more likely to undergo fracture perpendicular to those directions. Therefore, the present simulations suggest that fracture perpendicular to the $x$ and $y$ axes is more probable because of the magnitude of the corresponding residual stresses $\boldsymbol{\sigma}_{xx}$ and $\boldsymbol{\sigma}_{yy}$. These two directions have similar residual stresses, as shown in Figure \ref{fig:resstress}. This is consistent with laser-induced microcracking \cite{VRANCKEN2020464}, in which fracture is observed at the top surface perpendicular to the laser scan direction.

\section{Conclusions}
\label{sec:conclusions}

A modelling approach to predict residual stresses at the grain scale in additive manufactured 316 stainless steel is developed.
Thermal-fluid flow and grain growth simulations are used to provide the temperature field and the grain structure respectively, which are implemented in crystal plasticity finite element simulations. A method for element elimination and reactivation is developed to simulate melting and solidification during the laser scan. This method allows to reinitialise state variables, such as the plastic deformation, when elements are reactivated and it represents a step forward compared with previous studies. \\
Simulations show that residual stresses are strongly correlated with the plastic deformation, which in turn depends on the grain orientation. The directions with largest residual stresses are on the laser scan plane, both parallel and perpendicular to the laser scan direction. This is consistent with the microcracks observed experimentally. Even though this study is focused on 316 stainless steel and selective laser melting, the computational method developed can be applied to other metals and other additive manufacturing processes like directed energy deposition and electron beam additive manufacturing. More detailed simulations of residual stresses will be conducted in the future with experimental validation to systematically investigate the relationships between the manufacturing parameters, molten pool flow, grain structures, and residual stresses.

\section*{Acknowledgement}
Thanks to Dr. Yin Zhang from Georgia Institute of Technology for providing EBSD data used to obtain the Euler angles for grain orientations. Thanks to Prof. Yinmin Morris Wang for useful discussion on his experiments on AM 316L stainless steel. This research is supported by the Ministry of Education, Singapore, under its Academic Research Fund Tier 2 (MOE-T2EP50120-0012).

\clearpage

\bibliographystyle{model1-num-names}
\bibliography{references.bib}

\begin{thebibliography}{81}
\expandafter\ifx\csname natexlab\endcsname\relax\def\natexlab#1{#1}\fi
\providecommand{\bibinfo}[2]{#2}
\ifx\xfnm\relax \def\xfnm[#1]{\unskip,\space#1}\fi
%Type = Article
\bibitem[{Smith et~al.(2016)Smith, Xiong, Yan, Lin, Cheng, Kafka, Wagner, Cao,
  and Liu}]{YanSmith2016}
\bibinfo{author}{J.~Smith}, \bibinfo{author}{W.~Xiong},
  \bibinfo{author}{W.~Yan}, \bibinfo{author}{S.~Lin},
  \bibinfo{author}{P.~Cheng}, \bibinfo{author}{O.~L. Kafka},
  \bibinfo{author}{G.~J. Wagner}, \bibinfo{author}{J.~Cao},
  \bibinfo{author}{W.~K. Liu},
\newblock \bibinfo{title}{Linking process, structure, property, and performance
  for metal-based additive manufacturing: computational approaches with
  experimental support},
\newblock \bibinfo{journal}{Computational Mechanics} \bibinfo{volume}{57}
  (\bibinfo{year}{2016}) \bibinfo{pages}{583--610}.
%Type = Article
\bibitem[{Schwerdtfeger and K{\"o}rner(2014)}]{schwerdtfeger2014selective}
\bibinfo{author}{J.~Schwerdtfeger}, \bibinfo{author}{C.~K{\"o}rner},
\newblock \bibinfo{title}{Selective electron beam melting of
  {T}i--48{A}l--2{N}b--2{C}r: Microstructure and aluminium loss},
\newblock \bibinfo{journal}{Intermetallics} \bibinfo{volume}{49}
  (\bibinfo{year}{2014}) \bibinfo{pages}{29--35}.
%Type = Article
\bibitem[{Kruth et~al.(2005)Kruth, Mercelis, Van~Vaerenbergh, Froyen, and
  Rombouts}]{kruth2005binding}
\bibinfo{author}{J.-P. Kruth}, \bibinfo{author}{P.~Mercelis},
  \bibinfo{author}{J.~Van~Vaerenbergh}, \bibinfo{author}{L.~Froyen},
  \bibinfo{author}{M.~Rombouts},
\newblock \bibinfo{title}{Binding mechanisms in selective laser sintering and
  selective laser melting},
\newblock \bibinfo{journal}{Rapid prototyping journal}  (\bibinfo{year}{2005}).
%Type = Article
\bibitem[{Paul et~al.(2014)Paul, Anand, and Gerner}]{paul2014effect}
\bibinfo{author}{R.~Paul}, \bibinfo{author}{S.~Anand},
  \bibinfo{author}{F.~Gerner},
\newblock \bibinfo{title}{Effect of thermal deformation on part errors in metal
  powder based additive manufacturing processes},
\newblock \bibinfo{journal}{Journal of manufacturing science and Engineering}
  \bibinfo{volume}{136} (\bibinfo{year}{2014}).
%Type = Article
\bibitem[{Li et~al.(2018)Li, Liu, Fang, and Guo}]{li2018residual}
\bibinfo{author}{C.~Li}, \bibinfo{author}{Z.~Liu}, \bibinfo{author}{X.~Fang},
  \bibinfo{author}{Y.~Guo},
\newblock \bibinfo{title}{Residual stress in metal additive manufacturing},
\newblock \bibinfo{journal}{Procedia Cirp} \bibinfo{volume}{71}
  (\bibinfo{year}{2018}) \bibinfo{pages}{348--353}.
%Type = Article
\bibitem[{Taheri et~al.(2017)Taheri, Shoaib, Koester, Bigelow, Collins, and
  Bond}]{taheri2017powder}
\bibinfo{author}{H.~Taheri}, \bibinfo{author}{M.~R. B.~M. Shoaib},
  \bibinfo{author}{L.~W. Koester}, \bibinfo{author}{T.~A. Bigelow},
  \bibinfo{author}{P.~C. Collins}, \bibinfo{author}{L.~J. Bond},
\newblock \bibinfo{title}{Powder-based additive manufacturing-a review of types
  of defects, generation mechanisms, detection, property evaluation and
  metrology},
\newblock \bibinfo{journal}{International Journal of Additive and Subtractive
  Materials Manufacturing} \bibinfo{volume}{1} (\bibinfo{year}{2017})
  \bibinfo{pages}{172--209}.
%Type = Article
\bibitem[{Xie et~al.(2020)Xie, Xiang, Du, Yan, Shen, Chen, Shu, and
  Fang}]{XIE2020102723}
\bibinfo{author}{C.~Xie}, \bibinfo{author}{M.~Xiang}, \bibinfo{author}{J.~Du},
  \bibinfo{author}{W.~Yan}, \bibinfo{author}{L.~Shen},
  \bibinfo{author}{J.~Chen}, \bibinfo{author}{X.~Shu},
  \bibinfo{author}{Q.~Fang},
\newblock \bibinfo{title}{Asymmetric yield effect evolving with internal
  variables during continuous large deformations and its fem validation},
\newblock \bibinfo{journal}{International Journal of Plasticity}
  \bibinfo{volume}{130} (\bibinfo{year}{2020}) \bibinfo{pages}{102723}.
%Type = Article
\bibitem[{Mukherjee et~al.(2017)Mukherjee, Zhang, and
  DebRoy}]{MUKHERJEE2017360}
\bibinfo{author}{T.~Mukherjee}, \bibinfo{author}{W.~Zhang},
  \bibinfo{author}{T.~DebRoy},
\newblock \bibinfo{title}{An improved prediction of residual stresses and
  distortion in additive manufacturing},
\newblock \bibinfo{journal}{Computational Materials Science}
  \bibinfo{volume}{126} (\bibinfo{year}{2017}) \bibinfo{pages}{360--372}.
%Type = Article
\bibitem[{Mori(2006)}]{MORI20066737}
\bibinfo{author}{K.~Mori},
\newblock \bibinfo{title}{Finite element simulation of powder forming and
  sintering},
\newblock \bibinfo{journal}{Computer Methods in Applied Mechanics and
  Engineering} \bibinfo{volume}{195} (\bibinfo{year}{2006})
  \bibinfo{pages}{6737--6749}. \bibinfo{note}{Computational Metal Forming}.
%Type = Article
\bibitem[{Grilli et~al.(2021{\natexlab{a}})Grilli, Tarleton, and
  Cocks}]{GrilliFracturePF2021}
\bibinfo{author}{N.~Grilli}, \bibinfo{author}{E.~Tarleton},
  \bibinfo{author}{A.~C. Cocks},
\newblock \bibinfo{title}{Coupling a discrete twin model with cohesive elements
  to understand twin-induced fracture},
\newblock \bibinfo{journal}{International Journal of Fracture}
  \bibinfo{volume}{227} (\bibinfo{year}{2021}{\natexlab{a}})
  \bibinfo{pages}{173 -- 192}.
%Type = Article
\bibitem[{Grilli et~al.(2021{\natexlab{b}})Grilli, Tarleton, and
  Cocks}]{GRILLI2021100651}
\bibinfo{author}{N.~Grilli}, \bibinfo{author}{E.~Tarleton},
  \bibinfo{author}{A.~C. Cocks},
\newblock \bibinfo{title}{Neper2{CAE} and {P}y{C}i{G}en: Scripts to generate
  polycrystals and interface elements in abaqus},
\newblock \bibinfo{journal}{SoftwareX} \bibinfo{volume}{13}
  (\bibinfo{year}{2021}{\natexlab{b}}) \bibinfo{pages}{100651}.
%Type = Article
\bibitem[{Grilli et~al.(2020)Grilli, Cocks, and Tarleton}]{grilli2020crystal}
\bibinfo{author}{N.~Grilli}, \bibinfo{author}{A.~C. Cocks},
  \bibinfo{author}{E.~Tarleton},
\newblock \bibinfo{title}{Crystal plasticity finite element modelling of
  coarse-grained $\alpha$-uranium},
\newblock \bibinfo{journal}{Computational Materials Science}
  \bibinfo{volume}{171} (\bibinfo{year}{2020}) \bibinfo{pages}{109276}.
%Type = Inproceedings
\bibitem[{Grilli et~al.(2019)Grilli, Cocks, and Tarleton}]{GrilliCOMPLAS2019}
\bibinfo{author}{N.~Grilli}, \bibinfo{author}{A.~Cocks},
  \bibinfo{author}{E.~Tarleton},
\newblock \bibinfo{title}{Crystal plasticity finite element simulations of cast
  $\alpha$-uranium},
\newblock in: \bibinfo{editor}{E.~Onate}, \bibinfo{editor}{D.~Owen},
  \bibinfo{editor}{D.~Peric}, \bibinfo{editor}{M.~Chiumenti} (Eds.),
  \bibinfo{booktitle}{Computational plasticity XV: fundamentals and
  applications}. \bibinfo{note}{15th International Conference on Computational
  Plasticity - Fundamentals and Applications (COMPLAS), Barcelona, Spain, Sep
  03-05, 2019}.
%Type = Article
\bibitem[{Hocine et~al.(2020)Hocine, {Van Swygenhoven}, {Van Petegem}, Chang,
  Maimaitiyili, Tinti, {Ferreira Sanchez}, Grolimund, and
  Casati}]{HOCINE202030}
\bibinfo{author}{S.~Hocine}, \bibinfo{author}{H.~{Van Swygenhoven}},
  \bibinfo{author}{S.~{Van Petegem}}, \bibinfo{author}{C.~S.~T. Chang},
  \bibinfo{author}{T.~Maimaitiyili}, \bibinfo{author}{G.~Tinti},
  \bibinfo{author}{D.~{Ferreira Sanchez}}, \bibinfo{author}{D.~Grolimund},
  \bibinfo{author}{N.~Casati},
\newblock \bibinfo{title}{Operando x-ray diffraction during laser 3d printing},
\newblock \bibinfo{journal}{Materials Today} \bibinfo{volume}{34}
  (\bibinfo{year}{2020}) \bibinfo{pages}{30--40}.
%Type = Article
\bibitem[{Vrancken et~al.(2014)Vrancken, Cain, Knutsen, and
  Van~Humbeeck}]{vrancken2014residual}
\bibinfo{author}{B.~Vrancken}, \bibinfo{author}{V.~Cain},
  \bibinfo{author}{R.~Knutsen}, \bibinfo{author}{J.~Van~Humbeeck},
\newblock \bibinfo{title}{Residual stress via the contour method in compact
  tension specimens produced via selective laser melting},
\newblock \bibinfo{journal}{Scripta Materialia} \bibinfo{volume}{87}
  (\bibinfo{year}{2014}) \bibinfo{pages}{29--32}.
%Type = Article
\bibitem[{Mathar et~al.(1934)}]{mathar1934determination}
\bibinfo{author}{J.~Mathar}, et~al.,
\newblock \bibinfo{title}{Determination of initial stresses by measuring the
  deformation around drilled holes},
\newblock \bibinfo{journal}{Trans. ASME} \bibinfo{volume}{56}
  (\bibinfo{year}{1934}) \bibinfo{pages}{249--254}.
%Type = Article
\bibitem[{Mart{\'\i}nez-Garc{\'\i}a et~al.(2019)Mart{\'\i}nez-Garc{\'\i}a,
  Pedrini, Weidmann, Killinger, Gadow, Osten, and Schmauder}]{martinez2019non}
\bibinfo{author}{V.~Mart{\'\i}nez-Garc{\'\i}a}, \bibinfo{author}{G.~Pedrini},
  \bibinfo{author}{P.~Weidmann}, \bibinfo{author}{A.~Killinger},
  \bibinfo{author}{R.~Gadow}, \bibinfo{author}{W.~Osten},
  \bibinfo{author}{S.~Schmauder},
\newblock \bibinfo{title}{Non-contact residual stress analysis method with
  displacement measurements in the nanometric range by laser made material
  removal and slm based beam conditioning on ceramic coatings},
\newblock \bibinfo{journal}{Surface and Coatings Technology}
  \bibinfo{volume}{371} (\bibinfo{year}{2019}) \bibinfo{pages}{14--19}.
%Type = Article
\bibitem[{Ganeriwala et~al.(2019)Ganeriwala, Strantza, King, Clausen, Phan,
  Levine, Brown, and Hodge}]{ganeriwala2019evaluation}
\bibinfo{author}{R.~Ganeriwala}, \bibinfo{author}{M.~Strantza},
  \bibinfo{author}{W.~King}, \bibinfo{author}{B.~Clausen},
  \bibinfo{author}{T.~Q. Phan}, \bibinfo{author}{L.~E. Levine},
  \bibinfo{author}{D.~W. Brown}, \bibinfo{author}{N.~Hodge},
\newblock \bibinfo{title}{Evaluation of a thermomechanical model for prediction
  of residual stress during laser powder bed fusion of {T}i-6{A}l-4{V}},
\newblock \bibinfo{journal}{Additive Manufacturing} \bibinfo{volume}{27}
  (\bibinfo{year}{2019}) \bibinfo{pages}{489--502}.
%Type = Article
\bibitem[{Liang et~al.(2018)Liang, Cheng, Chen, Yang, and
  To}]{liang2018modified}
\bibinfo{author}{X.~Liang}, \bibinfo{author}{L.~Cheng},
  \bibinfo{author}{Q.~Chen}, \bibinfo{author}{Q.~Yang}, \bibinfo{author}{A.~C.
  To},
\newblock \bibinfo{title}{A modified method for estimating inherent strains
  from detailed process simulation for fast residual distortion prediction of
  single-walled structures fabricated by directed energy deposition},
\newblock \bibinfo{journal}{Additive Manufacturing} \bibinfo{volume}{23}
  (\bibinfo{year}{2018}) \bibinfo{pages}{471--486}.
%Type = Article
\bibitem[{Schoinochoritis et~al.(2017)Schoinochoritis, Chantzis, and
  Salonitis}]{schoinochoritis2017simulation}
\bibinfo{author}{B.~Schoinochoritis}, \bibinfo{author}{D.~Chantzis},
  \bibinfo{author}{K.~Salonitis},
\newblock \bibinfo{title}{Simulation of metallic powder bed additive
  manufacturing processes with the finite element method: A critical review},
\newblock \bibinfo{journal}{Proceedings of the Institution of Mechanical
  Engineers, Part B: Journal of Engineering Manufacture} \bibinfo{volume}{231}
  (\bibinfo{year}{2017}) \bibinfo{pages}{96--117}.
%Type = Article
\bibitem[{Qian et~al.(2014)Qian, González-Albuixech, and
  Niffenegger}]{QIAN2014312}
\bibinfo{author}{G.~Qian}, \bibinfo{author}{V.~González-Albuixech},
  \bibinfo{author}{M.~Niffenegger},
\newblock \bibinfo{title}{Probabilistic assessment of a reactor pressure vessel
  subjected to pressurized thermal shocks by using crack distributions},
\newblock \bibinfo{journal}{Nuclear Engineering and Design}
  \bibinfo{volume}{270} (\bibinfo{year}{2014}) \bibinfo{pages}{312--324}.
%Type = Article
\bibitem[{Yan et~al.(2020)Yan, Lu, Jones, Yang, Fox, Witherell, Wagner, and
  Liu}]{yan2020data}
\bibinfo{author}{W.~Yan}, \bibinfo{author}{Y.~Lu}, \bibinfo{author}{K.~Jones},
  \bibinfo{author}{Z.~Yang}, \bibinfo{author}{J.~Fox},
  \bibinfo{author}{P.~Witherell}, \bibinfo{author}{G.~Wagner},
  \bibinfo{author}{W.~K. Liu},
\newblock \bibinfo{title}{Data-driven characterization of thermal models for
  powder-bed-fusion additive manufacturing},
\newblock \bibinfo{journal}{Additive Manufacturing} \bibinfo{volume}{36}
  (\bibinfo{year}{2020}) \bibinfo{pages}{101503}.
%Type = Article
\bibitem[{Bailey et~al.(2017)Bailey, Katinas, and Shin}]{bailey2017laser}
\bibinfo{author}{N.~S. Bailey}, \bibinfo{author}{C.~Katinas},
  \bibinfo{author}{Y.~C. Shin},
\newblock \bibinfo{title}{Laser direct deposition of aisi h13 tool steel powder
  with numerical modeling of solid phase transformation, hardness, and residual
  stresses},
\newblock \bibinfo{journal}{Journal of Materials Processing Technology}
  \bibinfo{volume}{247} (\bibinfo{year}{2017}) \bibinfo{pages}{223--233}.
%Type = Article
\bibitem[{Cheon et~al.(2016)Cheon, Kiran, and Na}]{cheon2016thermal}
\bibinfo{author}{J.~Cheon}, \bibinfo{author}{D.~V. Kiran},
  \bibinfo{author}{S.-J. Na},
\newblock \bibinfo{title}{Thermal metallurgical analysis of {GMA} welded
  {A}{H}36 steel using {CFD}--{FEM} framework},
\newblock \bibinfo{journal}{Materials \& Design} \bibinfo{volume}{91}
  (\bibinfo{year}{2016}) \bibinfo{pages}{230--241}.
%Type = Article
\bibitem[{Chen and Yan(2020)}]{chen2020high}
\bibinfo{author}{F.~Chen}, \bibinfo{author}{W.~Yan},
\newblock \bibinfo{title}{High-fidelity modelling of thermal stress for
  additive manufacturing by linking thermal-fluid and mechanical models},
\newblock \bibinfo{journal}{Materials \& Design} \bibinfo{volume}{196}
  (\bibinfo{year}{2020}) \bibinfo{pages}{109185}.
%Type = Article
\bibitem[{Yan et~al.(2018)Yan, Qian, Ge, Lin, Liu, Lin, and
  Wagner}]{YAN2018210}
\bibinfo{author}{W.~Yan}, \bibinfo{author}{Y.~Qian}, \bibinfo{author}{W.~Ge},
  \bibinfo{author}{S.~Lin}, \bibinfo{author}{W.~K. Liu},
  \bibinfo{author}{F.~Lin}, \bibinfo{author}{G.~J. Wagner},
\newblock \bibinfo{title}{Meso-scale modeling of multiple-layer fabrication
  process in selective electron beam melting: Inter-layer/track voids
  formation},
\newblock \bibinfo{journal}{Materials \& Design} \bibinfo{volume}{141}
  (\bibinfo{year}{2018}) \bibinfo{pages}{210--219}.
%Type = Article
\bibitem[{Wei et~al.(2021)Wei, Mukherjee, Zhang, Zuback, Knapp, De, and
  DebRoy}]{WEI2021100703}
\bibinfo{author}{H.~Wei}, \bibinfo{author}{T.~Mukherjee},
  \bibinfo{author}{W.~Zhang}, \bibinfo{author}{J.~Zuback},
  \bibinfo{author}{G.~Knapp}, \bibinfo{author}{A.~De},
  \bibinfo{author}{T.~DebRoy},
\newblock \bibinfo{title}{Mechanistic models for additive manufacturing of
  metallic components},
\newblock \bibinfo{journal}{Progress in Materials Science}
  \bibinfo{volume}{116} (\bibinfo{year}{2021}) \bibinfo{pages}{100703}.
%Type = Article
\bibitem[{Stender et~al.(2018)Stender, Beghini, Sugar, Veilleux, Subia, Smith,
  San~Marchi, Brown, and Dagel}]{stender2018thermal}
\bibinfo{author}{M.~E. Stender}, \bibinfo{author}{L.~L. Beghini},
  \bibinfo{author}{J.~D. Sugar}, \bibinfo{author}{M.~G. Veilleux},
  \bibinfo{author}{S.~R. Subia}, \bibinfo{author}{T.~R. Smith},
  \bibinfo{author}{C.~W. San~Marchi}, \bibinfo{author}{A.~A. Brown},
  \bibinfo{author}{D.~J. Dagel},
\newblock \bibinfo{title}{A thermal-mechanical finite element workflow for
  directed energy deposition additive manufacturing process modeling},
\newblock \bibinfo{journal}{Additive Manufacturing} \bibinfo{volume}{21}
  (\bibinfo{year}{2018}) \bibinfo{pages}{556--566}.
%Type = Article
\bibitem[{Montevecchi et~al.(2016)Montevecchi, Venturini, Scippa, and
  Campatelli}]{montevecchi2016finite}
\bibinfo{author}{F.~Montevecchi}, \bibinfo{author}{G.~Venturini},
  \bibinfo{author}{A.~Scippa}, \bibinfo{author}{G.~Campatelli},
\newblock \bibinfo{title}{Finite element modelling of
  wire-arc-additive-manufacturing process},
\newblock \bibinfo{journal}{Procedia Cirp} \bibinfo{volume}{55}
  (\bibinfo{year}{2016}) \bibinfo{pages}{109--114}.
%Type = Article
\bibitem[{Yang et~al.(2016)Yang, Zhang, Cheng, Min, Chyu, and
  To}]{yang2016finite}
\bibinfo{author}{Q.~Yang}, \bibinfo{author}{P.~Zhang},
  \bibinfo{author}{L.~Cheng}, \bibinfo{author}{Z.~Min},
  \bibinfo{author}{M.~Chyu}, \bibinfo{author}{A.~C. To},
\newblock \bibinfo{title}{Finite element modeling and validation of
  thermomechanical behavior of {T}i-6{A}l-4{V} in directed energy deposition
  additive manufacturing},
\newblock \bibinfo{journal}{Additive Manufacturing} \bibinfo{volume}{12}
  (\bibinfo{year}{2016}) \bibinfo{pages}{169--177}.
%Type = Article
\bibitem[{Lindgren et~al.(1999)Lindgren, Runnemalm, and
  N{\"a}sstr{\"o}m}]{lindgren1999simulation}
\bibinfo{author}{L.-E. Lindgren}, \bibinfo{author}{H.~Runnemalm},
  \bibinfo{author}{M.~O. N{\"a}sstr{\"o}m},
\newblock \bibinfo{title}{Simulation of multipass welding of a thick plate},
\newblock \bibinfo{journal}{International journal for numerical methods in
  engineering} \bibinfo{volume}{44} (\bibinfo{year}{1999})
  \bibinfo{pages}{1301--1316}.
%Type = Article
\bibitem[{Lindgren and Hedblom(2001)}]{lindgren2001modelling}
\bibinfo{author}{L.-E. Lindgren}, \bibinfo{author}{E.~Hedblom},
\newblock \bibinfo{title}{Modelling of addition of filler material in large
  deformation analysis of multipass welding},
\newblock \bibinfo{journal}{Communications in numerical methods in engineering}
  \bibinfo{volume}{17} (\bibinfo{year}{2001}) \bibinfo{pages}{647--657}.
%Type = Article
\bibitem[{Michaleris(2014)}]{michaleris2014modeling}
\bibinfo{author}{P.~Michaleris},
\newblock \bibinfo{title}{Modeling metal deposition in heat transfer analyses
  of additive manufacturing processes},
\newblock \bibinfo{journal}{Finite Elements in Analysis and Design}
  \bibinfo{volume}{86} (\bibinfo{year}{2014}) \bibinfo{pages}{51--60}.
%Type = Phdthesis
\bibitem[{Ales(2018)}]{ales2018integrated}
\bibinfo{author}{T.~K. Ales}, \bibinfo{title}{An integrated model for the
  probabilistic prediction of yield strength in electron-beam additively
  manufactured {T}i-6{A}l-4{V}}, Ph.D. thesis, Iowa State University,
  \bibinfo{year}{2018}.
%Type = Article
\bibitem[{Miehe et~al.(2010)Miehe, Welschinger, and Hofacker}]{Miehe2010}
\bibinfo{author}{C.~Miehe}, \bibinfo{author}{F.~Welschinger},
  \bibinfo{author}{M.~Hofacker},
\newblock \bibinfo{title}{Thermodynamically consistent phase-field models of
  fracture: Variational principles and multi-field fe implementations},
\newblock \bibinfo{journal}{International Journal for Numerical Methods in
  Engineering} \bibinfo{volume}{83} (\bibinfo{year}{2010})
  \bibinfo{pages}{1273--1311}.
%Type = Book
\bibitem[{Borden et~al.(2018)Borden, Hughes, Landis, Anvari, and
  Lee}]{borden2018}
\bibinfo{author}{M.~Borden}, \bibinfo{author}{T.~Hughes},
  \bibinfo{author}{C.~Landis}, \bibinfo{author}{A.~Anvari},
  \bibinfo{author}{I.~Lee}, \bibinfo{title}{Phase-Field Formulation for Ductile
  Fracture}, \bibinfo{publisher}{In: Oñate E., Peric D., de Souza Neto E.,
  Chiumenti M. (eds) Advances in Computational Plasticity. Computational
  Methods in Applied Sciences, vol 46. Springer, Cham.}, \bibinfo{year}{2018}.
%Type = Article
\bibitem[{Grilli et~al.(2018)Grilli, Duarte, and Koslowski}]{GrilliDuarte2018}
\bibinfo{author}{N.~Grilli}, \bibinfo{author}{C.~A. Duarte},
  \bibinfo{author}{M.~Koslowski},
\newblock \bibinfo{title}{Dynamic fracture and hot-spot modeling in energetic
  composites},
\newblock \bibinfo{journal}{Journal of Applied Physics} \bibinfo{volume}{123}
  (\bibinfo{year}{2018}) \bibinfo{pages}{065101}.
%Type = Article
\bibitem[{Duarte et~al.(2018)Duarte, Grilli, and Koslowski}]{DuarteGrilli2018}
\bibinfo{author}{C.~A. Duarte}, \bibinfo{author}{N.~Grilli},
  \bibinfo{author}{M.~Koslowski},
\newblock \bibinfo{title}{Effect of initial damage variability on hot-spot
  nucleation in energetic materials},
\newblock \bibinfo{journal}{Journal of Applied Physics} \bibinfo{volume}{124}
  (\bibinfo{year}{2018}) \bibinfo{pages}{025104}.
%Type = Article
\bibitem[{Grilli and Koslowski(2019)}]{GrilliKoslowski2019}
\bibinfo{author}{N.~Grilli}, \bibinfo{author}{M.~Koslowski},
\newblock \bibinfo{title}{The effect of crystal anisotropy and plastic response
  on the dynamic fracture of energetic materials},
\newblock \bibinfo{journal}{Journal of Applied Physics} \bibinfo{volume}{126}
  (\bibinfo{year}{2019}) \bibinfo{pages}{155101}.
%Type = Article
\bibitem[{Chen et~al.(2019)Chen, Voisin, Zhang, Florien, Spadaccini, McDowell,
  Zhu, and Wang}]{chen2019microscale}
\bibinfo{author}{W.~Chen}, \bibinfo{author}{T.~Voisin},
  \bibinfo{author}{Y.~Zhang}, \bibinfo{author}{J.-B. Florien},
  \bibinfo{author}{C.~M. Spadaccini}, \bibinfo{author}{D.~L. McDowell},
  \bibinfo{author}{T.~Zhu}, \bibinfo{author}{Y.~M. Wang},
\newblock \bibinfo{title}{Microscale residual stresses in additively
  manufactured stainless steel},
\newblock \bibinfo{journal}{Nature communications} \bibinfo{volume}{10}
  (\bibinfo{year}{2019}) \bibinfo{pages}{1--12}.
%Type = Article
\bibitem[{Wang et~al.(2020)Wang, Ouyang, Fan, Guo, Li, Yan, and
  Li}]{yan2020MRL}
\bibinfo{author}{G.~Wang}, \bibinfo{author}{H.~Ouyang},
  \bibinfo{author}{C.~Fan}, \bibinfo{author}{Q.~Guo}, \bibinfo{author}{Z.~Li},
  \bibinfo{author}{W.~Yan}, \bibinfo{author}{Z.~Li},
\newblock \bibinfo{title}{The origin of high-density dislocations in additively
  manufactured metals},
\newblock \bibinfo{journal}{Materials Research Letters} \bibinfo{volume}{8}
  (\bibinfo{year}{2020}) \bibinfo{pages}{283--290}.
%Type = Article
\bibitem[{Grilli et~al.(2015)Grilli, Janssens, and {Van
  Swygenhoven}}]{GRILLI2015424}
\bibinfo{author}{N.~Grilli}, \bibinfo{author}{K.~G. Janssens},
  \bibinfo{author}{H.~{Van Swygenhoven}},
\newblock \bibinfo{title}{Crystal plasticity finite element modelling of low
  cycle fatigue in fcc metals},
\newblock \bibinfo{journal}{Journal of the Mechanics and Physics of Solids}
  \bibinfo{volume}{84} (\bibinfo{year}{2015}) \bibinfo{pages}{424--435}.
%Type = Article
\bibitem[{Grilli et~al.(2020)Grilli, Earp, Cocks, Marrow, and
  Tarleton}]{GrilliDIC2020}
\bibinfo{author}{N.~Grilli}, \bibinfo{author}{P.~Earp}, \bibinfo{author}{A.~C.
  Cocks}, \bibinfo{author}{J.~Marrow}, \bibinfo{author}{E.~Tarleton},
\newblock \bibinfo{title}{Characterisation of slip and twin activity using
  digital image correlation and crystal plasticity finite element simulation:
  Application to orthorhombic $\alpha$-uranium},
\newblock \bibinfo{journal}{Journal of the Mechanics and Physics of Solids}
  \bibinfo{volume}{135} (\bibinfo{year}{2020}) \bibinfo{pages}{103800}.
%Type = Article
\bibitem[{Yang et~al.(2021)Yang, Wang, and Yan}]{YangMin2021}
\bibinfo{author}{M.~Yang}, \bibinfo{author}{L.~Wang}, \bibinfo{author}{W.~Yan},
\newblock \bibinfo{title}{Phase-field modeling of grain evolutions in additive
  manufacturing from nucleation, growth, to coarsening},
\newblock \bibinfo{journal}{npj Computational Materials} \bibinfo{volume}{7}
  (\bibinfo{year}{2021}) \bibinfo{pages}{56}.
%Type = Article
\bibitem[{Yan et~al.(2018)Yan, Lian, Yu, Kafka, Liu, Liu, and
  Wagner}]{yan2018integrated}
\bibinfo{author}{W.~Yan}, \bibinfo{author}{Y.~Lian}, \bibinfo{author}{C.~Yu},
  \bibinfo{author}{O.~L. Kafka}, \bibinfo{author}{Z.~Liu},
  \bibinfo{author}{W.~K. Liu}, \bibinfo{author}{G.~J. Wagner},
\newblock \bibinfo{title}{An integrated process--structure--property modeling
  framework for additive manufacturing},
\newblock \bibinfo{journal}{Computer Methods in Applied Mechanics and
  Engineering} \bibinfo{volume}{339} (\bibinfo{year}{2018})
  \bibinfo{pages}{184--204}.
%Type = Article
\bibitem[{Asaro and Rice(1977)}]{asaro1977strain}
\bibinfo{author}{R.~J. Asaro}, \bibinfo{author}{J.~Rice},
\newblock \bibinfo{title}{Strain localization in ductile single crystals},
\newblock \bibinfo{journal}{Journal of the Mechanics and Physics of Solids}
  \bibinfo{volume}{25} (\bibinfo{year}{1977}) \bibinfo{pages}{309--338}.
%Type = Article
\bibitem[{Kalidindi(1998)}]{kalidindi1998incorporation}
\bibinfo{author}{S.~R. Kalidindi},
\newblock \bibinfo{title}{Incorporation of deformation twinning in crystal
  plasticity models},
\newblock \bibinfo{journal}{Journal of the Mechanics and Physics of Solids}
  \bibinfo{volume}{46} (\bibinfo{year}{1998}) \bibinfo{pages}{267--290}.
%Type = Article
\bibitem[{Vujo{\v{s}}evi{\'c} and Lubarda(2002)}]{vujovsevic2002finite}
\bibinfo{author}{L.~Vujo{\v{s}}evi{\'c}}, \bibinfo{author}{V.~Lubarda},
\newblock \bibinfo{title}{Finite-strain thermoelasticity based on
  multiplicative decomposition of deformation gradient},
\newblock \bibinfo{journal}{Theoretical and applied mechanics}
  (\bibinfo{year}{2002}) \bibinfo{pages}{379--399}.
%Type = Article
\bibitem[{Grilli et~al.(2020)Grilli, Cocks, and Tarleton}]{GRILLI2020104061}
\bibinfo{author}{N.~Grilli}, \bibinfo{author}{A.~C. Cocks},
  \bibinfo{author}{E.~Tarleton},
\newblock \bibinfo{title}{A phase field model for the growth and characteristic
  thickness of deformation-induced twins},
\newblock \bibinfo{journal}{Journal of the Mechanics and Physics of Solids}
  \bibinfo{volume}{143} (\bibinfo{year}{2020}) \bibinfo{pages}{104061}.
%Type = Article
\bibitem[{Roters et~al.(2010)Roters, Eisenlohr, Hantcherli, Tjahjanto, Bieler,
  and Raabe}]{roters2010overview}
\bibinfo{author}{F.~Roters}, \bibinfo{author}{P.~Eisenlohr},
  \bibinfo{author}{L.~Hantcherli}, \bibinfo{author}{D.~D. Tjahjanto},
  \bibinfo{author}{T.~R. Bieler}, \bibinfo{author}{D.~Raabe},
\newblock \bibinfo{title}{Overview of constitutive laws, kinematics,
  homogenization and multiscale methods in crystal plasticity finite-element
  modeling: Theory, experiments, applications},
\newblock \bibinfo{journal}{Acta Materialia} \bibinfo{volume}{58}
  (\bibinfo{year}{2010}) \bibinfo{pages}{1152--1211}.
%Type = Article
\bibitem[{Roters et~al.(2019)Roters, Diehl, Shanthraj, Eisenlohr, Reuber, Wong,
  Maiti, Ebrahimi, Hochrainer, Fabritius, Nikolov, Friák, Fujita, Grilli,
  Janssens, Jia, Kok, Ma, Meier, Werner, Stricker, Weygand, and
  Raabe}]{ROTERS2019420}
\bibinfo{author}{F.~Roters}, \bibinfo{author}{M.~Diehl},
  \bibinfo{author}{P.~Shanthraj}, \bibinfo{author}{P.~Eisenlohr},
  \bibinfo{author}{C.~Reuber}, \bibinfo{author}{S.~Wong},
  \bibinfo{author}{T.~Maiti}, \bibinfo{author}{A.~Ebrahimi},
  \bibinfo{author}{T.~Hochrainer}, \bibinfo{author}{H.-O. Fabritius},
  \bibinfo{author}{S.~Nikolov}, \bibinfo{author}{M.~Friák},
  \bibinfo{author}{N.~Fujita}, \bibinfo{author}{N.~Grilli},
  \bibinfo{author}{K.~Janssens}, \bibinfo{author}{N.~Jia},
  \bibinfo{author}{P.~Kok}, \bibinfo{author}{D.~Ma},
  \bibinfo{author}{F.~Meier}, \bibinfo{author}{E.~Werner},
  \bibinfo{author}{M.~Stricker}, \bibinfo{author}{D.~Weygand},
  \bibinfo{author}{D.~Raabe},
\newblock \bibinfo{title}{{DAMASK} – {T}he {D}\"{u}sseldorf {A}dvanced
  {M}aterial {S}imulation {K}it for modeling multi-physics crystal plasticity,
  thermal, and damage phenomena from the single crystal up to the component
  scale},
\newblock \bibinfo{journal}{Computational Materials Science}
  \bibinfo{volume}{158} (\bibinfo{year}{2019}) \bibinfo{pages}{420--478}.
%Type = Article
\bibitem[{Rice(1971)}]{rice1971inelastic}
\bibinfo{author}{J.~R. Rice},
\newblock \bibinfo{title}{Inelastic constitutive relations for solids: an
  internal-variable theory and its application to metal plasticity},
\newblock \bibinfo{journal}{Journal of the Mechanics and Physics of Solids}
  \bibinfo{volume}{19} (\bibinfo{year}{1971}) \bibinfo{pages}{433--455}.
%Type = Article
\bibitem[{Peirce et~al.(1983)Peirce, Asaro, and Needleman}]{peirce1983material}
\bibinfo{author}{D.~Peirce}, \bibinfo{author}{R.~J. Asaro},
  \bibinfo{author}{A.~Needleman},
\newblock \bibinfo{title}{Material rate dependence and localized deformation in
  crystalline solids},
\newblock \bibinfo{journal}{Acta metallurgica} \bibinfo{volume}{31}
  (\bibinfo{year}{1983}) \bibinfo{pages}{1951--1976}.
%Type = Article
\bibitem[{Daymond and Bouchard(2006)}]{daymond2006elastoplastic}
\bibinfo{author}{M.~Daymond}, \bibinfo{author}{P.~Bouchard},
\newblock \bibinfo{title}{Elastoplastic deformation of 316 stainless steel
  under tensile loading at elevated temperatures},
\newblock \bibinfo{journal}{Metallurgical and Materials transactions A}
  \bibinfo{volume}{37} (\bibinfo{year}{2006}) \bibinfo{pages}{1863--1873}.
%Type = Article
\bibitem[{Grilli et~al.(2020)Grilli, Tarleton, Edmondson, Gussev, and
  Cocks}]{ORNLpaper2020}
\bibinfo{author}{N.~Grilli}, \bibinfo{author}{E.~Tarleton},
  \bibinfo{author}{P.~D. Edmondson}, \bibinfo{author}{M.~N. Gussev},
  \bibinfo{author}{A.~C.~F. Cocks},
\newblock \bibinfo{title}{In situ measurement and modelling of the growth and
  length scale of twins in $\ensuremath{\alpha}$-uranium},
\newblock \bibinfo{journal}{Phys. Rev. Materials} \bibinfo{volume}{4}
  (\bibinfo{year}{2020}) \bibinfo{pages}{043605}.
%Type = Article
\bibitem[{Kalidindi and Anand(1992)}]{kalidindi1992approximate}
\bibinfo{author}{S.~R. Kalidindi}, \bibinfo{author}{L.~Anand},
\newblock \bibinfo{title}{An approximate procedure for predicting the evolution
  of crystallographic texture in bulk deformation processing of fcc metals},
\newblock \bibinfo{journal}{International journal of mechanical sciences}
  \bibinfo{volume}{34} (\bibinfo{year}{1992}) \bibinfo{pages}{309--329}.
%Type = Article
\bibitem[{Diehl et~al.(2017)Diehl, Wicke, Shanthraj, Roters, Brueckner-Foit,
  and Raabe}]{Diehl2017}
\bibinfo{author}{M.~Diehl}, \bibinfo{author}{M.~Wicke},
  \bibinfo{author}{P.~Shanthraj}, \bibinfo{author}{F.~Roters},
  \bibinfo{author}{A.~Brueckner-Foit}, \bibinfo{author}{D.~Raabe},
\newblock \bibinfo{title}{Coupled crystal plasticity–phase field fracture
  simulation study on damage evolution around a void: Pore shape versus
  crystallographic orientation},
\newblock \bibinfo{journal}{JOM} \bibinfo{volume}{69} (\bibinfo{year}{2017})
  \bibinfo{pages}{872--878}.
%Type = Article
\bibitem[{Irastorza-Landa et~al.(2016)Irastorza-Landa, {Van Swygenhoven}, {Van
  Petegem}, Grilli, Bollhalder, Brandstetter, and
  Grolimund}]{IRASTORZALANDA2016184}
\bibinfo{author}{A.~Irastorza-Landa}, \bibinfo{author}{H.~{Van Swygenhoven}},
  \bibinfo{author}{S.~{Van Petegem}}, \bibinfo{author}{N.~Grilli},
  \bibinfo{author}{A.~Bollhalder}, \bibinfo{author}{S.~Brandstetter},
  \bibinfo{author}{D.~Grolimund},
\newblock \bibinfo{title}{Following dislocation patterning during fatigue},
\newblock \bibinfo{journal}{Acta Materialia} \bibinfo{volume}{112}
  (\bibinfo{year}{2016}) \bibinfo{pages}{184--193}.
%Type = Misc
\bibitem[{Callen(1998)}]{callen1998thermodynamics}
\bibinfo{author}{H.~B. Callen}, \bibinfo{title}{Thermodynamics and an
  introduction to thermostatistics}, \bibinfo{year}{1998}.
%Type = Article
\bibitem[{Dubrovinsky(2002)}]{Dubrovinsky2002}
\bibinfo{author}{L.~Dubrovinsky},
\newblock \bibinfo{title}{{Thermal Expansion and Equation of State}},
\newblock \bibinfo{journal}{Encyclopedia of Materials: Science and Technology}
  (\bibinfo{year}{2002}) \bibinfo{pages}{1--4}.
%Type = Article
\bibitem[{Yadroitsev and Yadroitsava(2015)}]{yadroitsev2015evaluation}
\bibinfo{author}{I.~Yadroitsev}, \bibinfo{author}{I.~Yadroitsava},
\newblock \bibinfo{title}{Evaluation of residual stress in stainless steel 316l
  and ti6al4v samples produced by selective laser melting},
\newblock \bibinfo{journal}{Virtual and Physical Prototyping}
  \bibinfo{volume}{10} (\bibinfo{year}{2015}) \bibinfo{pages}{67--76}.
%Type = Book
\bibitem[{Grimvall(1999)}]{grimvall1999thermophysical}
\bibinfo{author}{G.~Grimvall}, \bibinfo{title}{Thermophysical properties of
  materials}, \bibinfo{publisher}{Elsevier}, \bibinfo{year}{1999}.
%Type = Article
\bibitem[{Jiang et~al.(2012)Jiang, Zhang, and Woo}]{jiang2012using}
\bibinfo{author}{W.~Jiang}, \bibinfo{author}{Y.~Zhang},
  \bibinfo{author}{W.~Woo},
\newblock \bibinfo{title}{Using heat sink technology to decrease residual
  stress in 316l stainless steel welding joint: Finite element simulation},
\newblock \bibinfo{journal}{International journal of pressure vessels and
  piping} \bibinfo{volume}{92} (\bibinfo{year}{2012}) \bibinfo{pages}{56--62}.
%Type = Article
\bibitem[{Grilli and Koslowski(2018)}]{grilli2018effect}
\bibinfo{author}{N.~Grilli}, \bibinfo{author}{M.~Koslowski},
\newblock \bibinfo{title}{The effect of crystal orientation on shock loading of
  single crystal energetic materials},
\newblock \bibinfo{journal}{Computational Materials Science}
  \bibinfo{volume}{155} (\bibinfo{year}{2018}) \bibinfo{pages}{235--245}.
%Type = Book
\bibitem[{Reed and Horiuchi(1982)}]{Horiuci316SS}
\bibinfo{author}{R.~Reed}, \bibinfo{author}{T.~Horiuchi},
  \bibinfo{title}{Austenitic steels at low temperatures},
  \bibinfo{publisher}{Plenum press}, \bibinfo{address}{New York},
  \bibinfo{year}{1982}.
%Type = Article
\bibitem[{Grilli et~al.(2018)Grilli, Janssens, Nellessen, Sandlöbes, and
  Raabe}]{GRILLI2018104}
\bibinfo{author}{N.~Grilli}, \bibinfo{author}{K.~Janssens},
  \bibinfo{author}{J.~Nellessen}, \bibinfo{author}{S.~Sandlöbes},
  \bibinfo{author}{D.~Raabe},
\newblock \bibinfo{title}{Multiple slip dislocation patterning in a
  dislocation-based crystal plasticity finite element method},
\newblock \bibinfo{journal}{International Journal of Plasticity}
  \bibinfo{volume}{100} (\bibinfo{year}{2018}) \bibinfo{pages}{104--121}.
%Type = Article
\bibitem[{Yan et~al.(2018)Yan, Lin, Kafka, Yu, Liu, Lian, Wolff, Cao, Wagner,
  and Liu}]{YanLin2018}
\bibinfo{author}{W.~Yan}, \bibinfo{author}{S.~Lin}, \bibinfo{author}{O.~L.
  Kafka}, \bibinfo{author}{C.~Yu}, \bibinfo{author}{Z.~Liu},
  \bibinfo{author}{Y.~Lian}, \bibinfo{author}{S.~Wolff},
  \bibinfo{author}{J.~Cao}, \bibinfo{author}{G.~J. Wagner},
  \bibinfo{author}{W.~K. Liu},
\newblock \bibinfo{title}{Modeling process-structure-property relationships for
  additive manufacturing},
\newblock \bibinfo{journal}{Frontiers of Mechanical Engineering}
  \bibinfo{volume}{13} (\bibinfo{year}{2018}) \bibinfo{pages}{482--492}.
%Type = Article
\bibitem[{Permann et~al.(2020)Permann, Gaston, Andr{\v{s}}, Carlsen, Kong,
  Lindsay, Miller, Peterson, Slaughter, Stogner, and
  Martineau}]{permann2020moose}
\bibinfo{author}{C.~J. Permann}, \bibinfo{author}{D.~R. Gaston},
  \bibinfo{author}{D.~Andr{\v{s}}}, \bibinfo{author}{R.~W. Carlsen},
  \bibinfo{author}{F.~Kong}, \bibinfo{author}{A.~D. Lindsay},
  \bibinfo{author}{J.~M. Miller}, \bibinfo{author}{J.~W. Peterson},
  \bibinfo{author}{A.~E. Slaughter}, \bibinfo{author}{R.~H. Stogner},
  \bibinfo{author}{R.~C. Martineau},
\newblock \bibinfo{title}{{MOOSE}: Enabling massively parallel multiphysics
  simulation},
\newblock \bibinfo{journal}{{SoftwareX}} \bibinfo{volume}{11}
  (\bibinfo{year}{2020}) \bibinfo{pages}{100430}.
%Type = Article
\bibitem[{Adhikary et~al.(2016)Adhikary, Jayasundara, Podgorney, and
  Wilkins}]{adhikary2016robust}
\bibinfo{author}{D.~P. Adhikary}, \bibinfo{author}{C.~Jayasundara},
  \bibinfo{author}{R.~K. Podgorney}, \bibinfo{author}{A.~H. Wilkins},
\newblock \bibinfo{title}{A robust return-map algorithm for general
  multisurface plasticity},
\newblock \bibinfo{journal}{International Journal for Numerical Methods in
  Engineering} \bibinfo{volume}{109} (\bibinfo{year}{2016})
  \bibinfo{pages}{218--234}.
%Type = Article
\bibitem[{Chockalingam et~al.(2013)Chockalingam, Tonks, Hales, Gaston, Millett,
  and Zhang}]{chockalingam2013crystal}
\bibinfo{author}{K.~Chockalingam}, \bibinfo{author}{M.~Tonks},
  \bibinfo{author}{J.~Hales}, \bibinfo{author}{D.~Gaston},
  \bibinfo{author}{P.~Millett}, \bibinfo{author}{L.~Zhang},
\newblock \bibinfo{title}{Crystal plasticity with jacobian-free
  newton--krylov},
\newblock \bibinfo{journal}{Computational Mechanics} \bibinfo{volume}{51}
  (\bibinfo{year}{2013}) \bibinfo{pages}{617--627}.
%Type = Article
\bibitem[{Lee(1969)}]{lee1969elastic}
\bibinfo{author}{E.~H. Lee},
\newblock \bibinfo{title}{Elastic-plastic deformation at finite strains}
  (\bibinfo{year}{1969}).
%Type = Article
\bibitem[{Irastorza-Landa et~al.(2017)Irastorza-Landa, Grilli, and
  Van~Swygenhoven}]{irastorza2017effect}
\bibinfo{author}{A.~Irastorza-Landa}, \bibinfo{author}{N.~Grilli},
  \bibinfo{author}{H.~Van~Swygenhoven},
\newblock \bibinfo{title}{Effect of pre-existing immobile dislocations on the
  evolution of geometrically necessary dislocations during fatigue},
\newblock \bibinfo{journal}{Modelling and Simulation in Materials Science and
  Engineering} \bibinfo{volume}{25} (\bibinfo{year}{2017})
  \bibinfo{pages}{055010}.
%Type = Article
\bibitem[{Clausen et~al.(1998)Clausen, Lorentzen, and Leffers}]{Clausen1998}
\bibinfo{author}{B.~Clausen}, \bibinfo{author}{T.~Lorentzen},
  \bibinfo{author}{T.~Leffers},
\newblock \bibinfo{title}{{Self-consistent modelling of the plastic deformation
  of F.C.C. polycrstals and its implications for diffraction measurements of
  internal stresses}},
\newblock \bibinfo{journal}{Acta Materialia} \bibinfo{volume}{46}
  (\bibinfo{year}{1998}) \bibinfo{pages}{3087--3098}.
%Type = Article
\bibitem[{Yushu et~al.(2021)Yushu, Jiang, Tan, Sun, Schwen, and
  Spencer}]{Yushu2021ElemElim}
\bibinfo{author}{D.~Yushu}, \bibinfo{author}{W.~Jiang},
  \bibinfo{author}{C.~Tan}, \bibinfo{author}{C.~Sun},
  \bibinfo{author}{D.~Schwen}, \bibinfo{author}{B.~Spencer},
\newblock \bibinfo{title}{A thermal-mechanical model for directed energy
  deposition process with element activation},
\newblock \bibinfo{journal}{(submitted manuscript)}  (\bibinfo{year}{2021}).
%Type = Article
\bibitem[{Duarte et~al.(2019)Duarte, Koslowski, and Grilli}]{Duarte2019APS}
\bibinfo{author}{C.~A. Duarte}, \bibinfo{author}{M.~Koslowski},
  \bibinfo{author}{N.~Grilli},
\newblock \bibinfo{title}{Modeling the effect of plasticity and damage in
  $\beta$-hmx single crystals under shock loading},
\newblock \bibinfo{journal}{21st Biennial Conference of the APS Topical Group
  on Shock Compression of Condensed Matter} \bibinfo{volume}{64}
  (\bibinfo{year}{2019}).
%Type = Article
\bibitem[{Yan et~al.(2017)Yan, Ge, Qian, Lin, Zhou, Liu, Lin, and
  Wagner}]{yan2017multi}
\bibinfo{author}{W.~Yan}, \bibinfo{author}{W.~Ge}, \bibinfo{author}{Y.~Qian},
  \bibinfo{author}{S.~Lin}, \bibinfo{author}{B.~Zhou}, \bibinfo{author}{W.~K.
  Liu}, \bibinfo{author}{F.~Lin}, \bibinfo{author}{G.~J. Wagner},
\newblock \bibinfo{title}{Multi-physics modeling of single/multiple-track
  defect mechanisms in electron beam selective melting},
\newblock \bibinfo{journal}{Acta Materialia} \bibinfo{volume}{134}
  (\bibinfo{year}{2017}) \bibinfo{pages}{324--333}.
%Type = Article
\bibitem[{Wang et~al.(2020)Wang, Zhang, and Yan}]{wang2020evaporation}
\bibinfo{author}{L.~Wang}, \bibinfo{author}{Y.~Zhang},
  \bibinfo{author}{W.~Yan},
\newblock \bibinfo{title}{Evaporation model for keyhole dynamics during
  additive manufacturing of metal},
\newblock \bibinfo{journal}{Physical Review Applied} \bibinfo{volume}{14}
  (\bibinfo{year}{2020}) \bibinfo{pages}{064039}.
%Type = Article
\bibitem[{Simmons et~al.(2000)Simmons, Shen, and Wang}]{simmons2000phase}
\bibinfo{author}{J.~Simmons}, \bibinfo{author}{C.~Shen},
  \bibinfo{author}{Y.~Wang},
\newblock \bibinfo{title}{Phase field modeling of simultaneous nucleation and
  growth by explicitly incorporating nucleation events},
\newblock \bibinfo{journal}{Scripta materialia} \bibinfo{volume}{43}
  (\bibinfo{year}{2000}) \bibinfo{pages}{935--942}.
%Type = Phdthesis
\bibitem[{Grilli(2016)}]{Grilli2016thesis}
\bibinfo{author}{N.~Grilli}, \bibinfo{title}{Physics-based constitutive
  modelling for crystal plasticity finite element computation of cyclic
  plasticity in fatigue}, Ph.D. thesis, \'Ecole Polytechnique F\'ed\'erale de
  Lausanne, \bibinfo{year}{2016}.
%Type = Article
\bibitem[{Irastorza-Landa et~al.(2017)Irastorza-Landa, Grilli, and {Van
  Swygenhoven}}]{IRASTORZALANDA2017157}
\bibinfo{author}{A.~Irastorza-Landa}, \bibinfo{author}{N.~Grilli},
  \bibinfo{author}{H.~{Van Swygenhoven}},
\newblock \bibinfo{title}{Laue micro-diffraction and crystal plasticity finite
  element simulations to reveal a vein structure in fatigued {C}u},
\newblock \bibinfo{journal}{Journal of the Mechanics and Physics of Solids}
  \bibinfo{volume}{104} (\bibinfo{year}{2017}) \bibinfo{pages}{157--171}.
%Type = Article
\bibitem[{Vrancken et~al.(2020)Vrancken, Ganeriwala, and
  Matthews}]{VRANCKEN2020464}
\bibinfo{author}{B.~Vrancken}, \bibinfo{author}{R.~K. Ganeriwala},
  \bibinfo{author}{M.~J. Matthews},
\newblock \bibinfo{title}{Analysis of laser-induced microcracking in tungsten
  under additive manufacturing conditions: Experiment and simulation},
\newblock \bibinfo{journal}{Acta Materialia} \bibinfo{volume}{194}
  (\bibinfo{year}{2020}) \bibinfo{pages}{464--472}.

\end{thebibliography}

\end{document}